\documentclass[a4paper,11pt]{article}
\usepackage{pos}
\usepackage{amsmath}
\usepackage{color}
\usepackage{cancel}
\usepackage{slashed}
\usepackage{tikz}

\newcommand{\beq}{\begin{equation}}
\newcommand{\eeq}{\end{equation}}
\newcommand{\beqa}{\begin{eqnarray}}
\newcommand{\eeqa}{\end{eqnarray}}

\newcommand{\DO}{D_\Omega}
\newcommand{\Dd}{D_\partial}
\newcommand{\DOi}{D_\Omega^{-1}}

\newcommand{\DOb}{D_{\bar{\Omega}}}
\newcommand{\Ddb}{D_{\bar{\partial}}}
\newcommand{\DObi}{D_{\bar{\Omega}}^{-1}}

\newcommand{\Pdb}{\mathbb{P}_{\bar{\partial}}}

\newcommand{\hDdb}{\hat D_{\bar{\partial}}}

\newcommand{\hPdb}{\hat{\mathbb{P}}_{\bar{\partial}}}

\title{Advances in algorithms for solvers and gauge generation}

\author*[a]{Peter Boyle}

\affiliation[a]{Brookhaven National Laboratory, Upton, New York}

\emailAdd{pboyle@bnl.gov}

\abstract{
  I review recent research and advances in algorithms for solvers and gauge generation, with an emphasis
  on practical algorithms for four dimensional simulations. Particular consideration is given to advances
  in multigrid solvers, fourier acceleration and field transformation approaches to accelerating
  evolution dynamics, and to parallel tempering approaches to solving the topological tunneling problem.
  Particular consideration is given to the interaction between rapidly evolving computer architecture and optimal
  algorithms that exploit these.
  In this conference, nascent machine learning algorithms were separately reviewed\cite{Kanwar}.
}

\FullConference{The 40th International Symposium on Lattice Field Theory (Lattice 2023)\\
July 31st - August 4th, 2023\\
Fermi National Accelerator Laboratory\\}

\begin{document}
\maketitle

\section{Introduction}

Efficient algorithms are central to research in lattice gauge theory carried out around the world.
Over several decades now gains from algorithmic research and innovation have been at very least
comparable to decades of exponential growth in computing power under Moore's law.

The generation of gauge configurations importance samples from up to $10^{10}$ degrees of freedom using
Markov chain Monte Carlo (MCMC), typically performed using some variant of Metropolis-Rosenbuth-Rosenbluth-Teller-Teller
algorithm\cite{Metropolis:1953am}. These algorithms notably emerging in high energy physics and are now used broadly across much of science. The first
computer MCMC simulations were programmed and performed by Ariana Rosenbluth and Augusta Teller on the ``MANIAC'' computer at Los Alamos,
in scientific computing work that predated the first programming languages and compilers.
It is unfortunate that four key authors are often dropped in the common usage names
``Metropolis'' or ``Metropolis-Hastings'' algorithms. The algorithms were revolutionary as they show how
to sample a high dimensional space from an arbitrary probability distribution.
New challenges are emerging however, as simulation parameters have moved on from the inclusion of
of Fermions and the reduction of quark masses to direct simulation at the physical. Now, further computing
power is largely invested in increasingly fine resolution simulations with the introduction of a new facet to
the dynamics of lattice simulations: covering a significantly greater dynamic range of length scales in a single simulations.
Both solver and gauge evolution algorithms suffer from \emph{critical slowing down} which necessarily arises as the continuum limit
is taken.

The structure of this review is as follows: a background review of the state of the art and and issues with contemporary lattice
algorithms will be given in section~\ref{sec:CSD} and \ref{sec:fermCSD}. Recent developments in the area of gauge
evolution algorithms are reviewed in section~\ref{sec:gauge}, and recent developments in the area of Fermion solver
algorithms are reviewed in section~\ref{sec:multigrid}.

Throughout, and following the brief of the organising committee, particular attention will be paid to the interplay between
algorithms and computer architecture. High end computing is increasingly dominated by Graphics Processor Units (GPUs)
and the recent trends in computer architecture can be summarised as follows:
\begin{itemize}
\itemsep=0em
\item Peak floating point performance is increasing rapidly, particularly for large matrix multiplication and reduced precision operations.
\item Hardware is highly parallel: large amounts of fine grain work are required for good performance.
\item Memory organisation is complex: small local memory (such as GPU memory) can be much faster than large globally addressable memory.
\item Intranode interconnects are quite efficient, particularly those between GPU's in the same computing node.
\item Internode interconnect is an increasing bottleneck for parallel calculation.
\end{itemize}
These trends present both challenges, but also opportunities for new hardware sympathetic algorithms designed to exploit these features.

\section{Gauge configuration sampling and critical slowing down}

\label{sec:CSD}
The most practical forms of simulations with dynamical Fermions make use of some form of \emph{hybrid Monte Carlo} (HMC)
algorithm developed in the late 1980's\cite{Duane:1987de}.
It is recently common for HMC to be used in machine learning, where is synonymously labelled ``Hamiltonian Monte Carlo''.
The path integral for Euclidean lattice field theory is augmented with an auxiliary Gaussian integral over additional ``momentum'' degrees
of freedom with the same dimension as the Lie algebra of the gauge group. These degrees of freedom are sample jointly
with the gauge fields, merely contributing an overall factor to the partition function.
The Metropolis proposal for joint sampling is composed of a Gaussian momentum draw, and an evolution in a fictitious ``Monte Carlo time''
at (up to integration errors) constant joint probability. Compared to say, Langevin algorithms\cite{Horowitz:1985kd},
this allows for large(r) changes in the gauge field to be made with good acceptance probability, while the Metropolis accept/reject
step, along with use of a reversible and area preserving molecular dynamics
integration algorithm, ensures the algorithm is exact at non-zero integrator step size.

The Grassman Gaussian integral for Fermions yields the determinant of Dirac operator, and for two degenerate flavours this
may be replaced by a complex Gaussian integral of the inverse of the squared operator, and is then amenable to numerical integration.
Thus the QCD partition function involving a gauge action $S_G$ and two flavour pseudofermion for Dirac operator $M$ can be cast as,
\beq
  Z=
  \int d\pi
  \int d \phi
  \int d U \quad
  e^{-\frac{\pi^2}{2}}
  e^{-S_G[U]}
  e^{-\phi^\ast (M^\dag M)^{-1} \phi }
  =
  \int d\pi
  \int d \phi
  \int d U \quad
  e^{-\frac{\pi^2}{2}}
  e^{-S_{QCD}[U,\phi]}.
\eeq
The HMC Markov transition consists of,
  \begin{enumerate}
  \item Draw gaussian momenta $\pi$ and pseudofermion $\phi$ as gaussian $\eta = M^{-1} \phi$
  \item Inner \emph{molecular dynamics integration} at constant Hamiltonian (i.e. log likelihood):
\beq
  H= \frac{\pi^2}{2}+S_G[U]+\phi^\ast (M^\dag M)^{-1} \phi.
\eeq
  \item Metropolis acceptance-reject step.
  \end{enumerate}
  Detailed balance is assured if the molecular dynamics integrator is symplectic (area preserving) and reversible. If the
  process is also ergodic the target distribution becomes the asymptotic fixed point of the Markov process.
  Importantly, while the tail of Gaussian distribution makes the process formally ergodic, this may not practically be the case in
  a bounded amount of simulation time.
  The molecular dynamics equations of motion under $\dot U = i \pi U$, are easily derived via:
\beq
  \dot{H} = 0 = \pi \left[ \dot \pi + i  U \cdot \left(\nabla_U S_{QCD}\right)_{\rm TA} \right],
\eeq
  Where ${\rm TA}$ means the traceless anti-hermitian projection, and a finite timestep is performed via
  $U(t+dt) = e^{i \pi dt} U(t)$.
  The force terms $\nabla_U S_{QCD}$ require the inversion of $M^\dagger M$ at each timestep of evolution to evaluate the molecular dynamics force.
  Odd numbers of Fermion flavors can be handled by rational approximation via the rational hybrid
  Monte Carlo algorithm\cite{Clark:2003na,Clark:2004cp,Clark:2006wq}, 
  albeit 
  requiring positivity of the real eigenvalues which may not always be satisfied for certain lattice Fermion actions\cite{Mohler:2020txx}.
  In the case of domain wall Fermions an exact one flavour algorithm\cite{Chen:2014hyy,Chen:2017gsk} may be used.
  Exact algorithms are advantageous\cite{Clark:2005sq}, but bring a penalty that the cost to maintain a good acceptance rate increases
  with lattice volume. For a fairly typical modern lattice of $96^3\times 192$ there are O($10^{10}$) degrees of freedom,
  and we require to conserve the HMC Hamiltonian to an accuracy of as much as one part in $10^{10}$. 
  \label{sec:dH}
  This adverse volume scaling leads to several schools of thought on the asymptotic volume dependence of HMC,
  with variously advocacy of inexact algorithms\cite{Gottlieb:1987mq} which formally need a step size extrapolation,
  and also the use of stochastic molecular dynamics to reduce sensitivity to force spikes\cite{Francis:2019muy} in large volumes.
  However, the vast majority of contemporary QCD simulation is performed with some variant of HMC at this time.
  
  \subsection{Determinant factorisation}
  The Fermion determinant is estimated stochastically, and different estimates of the Fermion force can be more or less
  efficient, yielding different costs and sizes of noise components.
A significant element of cost reduction for dynamical fermions has involved various forms of \emph{determinant factorisation}
into multiple (and ideally computationally cheaper) factors whose product reconstructs the full determinant.
The most common of these is \emph{Hasenbusch mass preconditioning}\cite{Hasenbusch:2001ne,Urbach:2005ji}, replacing
a single two flavour pseudofermion determinant at a light quark mass with multiple determinant ratios with
intermediate, and gradually increased quark masses and which are each treated with independent pseudofermion
integrals.
$$
\det D^\dagger D(m_{ud}) = \det\frac{ D^\dagger D(m_{ud})}{ D^\dagger D(m_{h})} \det D^\dagger D(m_{h})
$$
The forces for nearby masses are reduced, and in the case of
a Fermion action with an additive mass term clearly probing only the lowest eigenvalue modes of the system.
The bulk of the fermion force is in the denser higher eigenvalue region while, the multiple larger quark masses used are relatively
cheap to compute.

A second, common form of determinant splitting is used in domain decomposed HMC (DDHMC)\cite{Luscher:2004pav,DelDebbio:2006cn,DelDebbio:2007pz}: namely coordinate space Schur factorisation into subdomain cells.
DDHMC encouragingly imposes locality on much of the pseudofermion action with boundary determinants of a Schur complement.
A Fermion operator may be factored around hyper-cuboidal cells partitioning the lattice into two colors of cell in a checkerboarded fashion:
\begin{equation}
 \left(\begin{array}{cc} \DO & \Dd \\
                            \Ddb&\DOb \end{array}
   \right)
= 
\left(
\begin{array}{cc}
1  &  \Dd \DObi \\
0  & 1
\end{array}
\right)
\left(
\begin{array}{cc}
\DO - \Dd \DObi \Ddb & 0\\
0                    & \DOb
\end{array}
\right)
\left(
\begin{array}{cc}
1 &  0 \\
\DObi \Ddb  & 1
\end{array}
\right).
\end{equation}
$$
\det D = \det{\DO} \det \DOb \det \left\{ 1 - \DOi \Dd \DObi \Ddb\right\},
$$
The subdomain determinants are purely local, maintaining Dirichlet boundary conditions at the cell boundaries
while the boundary determinant can be integrated on a coarser timestep if the forces can be kept bounded.
The forces may kept limited, to a degree, by spatial separation of ``active'' links from the boundary.

Domain decomposition is attractive in two ways. From a computing perspective breaks the pseudofermion action into
strictly local component that involves no communication between domains and ring fences the non-locality in a boundary
determinant whose force sizes are in principle controllable by physically separating updated links whose ``force'' is
evaluated with respect to a determinant ratio that deviates from unity due only to terms at the boundary between domains.

A third form of determinant splitting is known as the n-roots trick\cite{Clark:2006fx}, where n-pseudofermions
are introduced each estimating the n-th root of the determinant by rational approximation (with a corresponding force reduction as this is closer
to unity) 
$$
\det D = \left( \det D^\frac{1}{n} \right)^n
$$
Each factor is not intrinsically of particularly lower cost, however, de Forcrand and Keegan recently observed with staggered Fermions
that the use of block solvers can reduce cost giving an overall benefit\cite{deForcrand:2018orx}.

\subsection{Critical slowing down of the HMC algorithms}

For hybrid Monte Carlo, the problem of critical slowing down can be easily seen in the case of free scalar fields\cite{Kennedy:1999di,Kennedy:2000ju}.
In gauge theories, in the free limit the transverse physical momentum modes behave
as decoupled harmonic oscillators with a period in molecular dynamics time that is dependent on the momentum
(angular wavenumber) $k$\cite{Sheta:2021hsd},
$$
\omega_k = \sqrt{\frac{\beta}{6}} k.
$$
Since the angular frequency in molecular dynamics time is strongly dependent on the virtuality of a mode, there is a large dynamic range
in the size of forces per mode, diverging in the ultraviolet and requiring both timestep reduction. Further 
there is \emph{no} common integration trajectory length that will simultaneously maximally decorrelate all lengthscales in the sampled field.
This gives rise to a generalised critical slowing down as the continuum limit is taken, associated with the mutiscale dynamics.

\subsubsection{Topological tunneling}

It is further the case that topologically indeterminate configurations will receive a diverging action penalty
as the continuum limit is taken due to the gauge action of a non-smooth configuation\cite{DelDebbio:2002xa,DelDebbio:2004xh}.
This has been observed empirically to result in topological freezing 
in practical simulations, where depending on the gauge and Fermion action simulations with inverse lattice
spacings between 3 and 4 GeV have been seen to become practically non-ergodic with \emph{frozen} topology,
and thus demonstrable non-ergodicity\cite{DelDebbio:2002xa,DelDebbio:2004xh,Luscher:2010we,Schaefer:2010hu}.
The interpretation and response to this fulfill a basic requirement for correctness of a sampling algorithm
has been varied and debatable largely because the only clean solution would be for it not to happen.

It has been recognised that changing the periodic gauge boundary conditions in typical lattice calculations
to being open in the time direction will allow the ends to ``flap'' and in principle for topological charge density to flow
in and out of the lattice\cite{Luscher:2011kk}.
However, this change of boundary condition also \emph{changes} the meaning of global topology on the lattice as a surface term.
While this may fix the symptom, it is possible that there is a generalised critical slowing down and that \emph{other}
observables may not be similarly be ``fixed''. The symptom as an indicator of pathology has been modified such that it can no longer
show the pathology: the cure addresses topological freezing by making the ends of the lattice ``warm'', but it does not
necessarily preclude the presence of other long autocorrelation modes that remain unaddressed.

There is some evidence\cite{McGlynn:2014bxa} that generalised slowing down of low-noise long distance observables may in fact
occur with Wilson flowed energy densities also slowing down. With open boundary conditions both local topological
charge activity and energy densities must be observed away from the modified boundary.
The scope for variance reduction from volume averaging in valence Fermionic correlation is somewhat impaired with open boundary conditions
as observables should not approach the boundaries, and the temporal extents increased. Especially in analyses where
eigenvector deflation and volume averaging of low modes are used (with an $O(V^2)$ cost), the increase in lattice volume can be prohibitive.

For these and other reasons, it would be more ideal to address topological freezing by algorithmic improvement
without modifying the boundary conditions. The problem of topology tunneling
may be sufficiently distinct in nature from a generalised critical slowing down of sampling within a topological
sector that algoritmic solutions may in fact be distinct and complementary.
New, and potentially complementary, ideas addressing topological sampling and fourier acceleration
will addressed in section~\ref{sec:PT} and section~\ref{sec:FA}.
This is a point that may be important to keep in mind when using or investigating critical slowing down in low dimensional
proxy models which have a topological index, but have simpler dynamics than QCD.

\label{Sec:ideal}

\section{Critical slowing down of Fermion solvers}

\label{sec:fermCSD}
In Fermion Krylov solver algorithms, critical slowing down is dictated by the domain in the complex plane overwhich
a Krylov solver polynomial must accurately reproduce the reciprocal of each eigenvalue of the matrix.
In the case of Hermitian solver algorithms,
critical slowing down is dictated by the condition number, $\kappa$, as the ratio of the highest to smallest eigenvalues.
Krylov solvers select the best solution (under some metric) in the space spanned by polynomials of the matrix applied to the
right hand side that is being solved, and the growth of this domain of approximation causes growth in the number of
solver iterations and loss of algorithmic efficiency.

In the case of conjugate gradient on the squared operator, a sketch of critical slowing down may
be performed for Wilson Fermions with a free field analysis, $\kappa=\frac{64}{a^2 m_{ud}}$. A minimax analysis based on Chebyshev
polynomials \cite{saad} leads to a worst case bound on the residual reduction per iteration of
\beq
\sigma = \frac{\sqrt{k}-1}{\sqrt{k}+1}.
\eeq
This bound will be saturated in the large volume limit where the spectrum becomes dense.
An estimate for the number of iterations required to solve to a given residual reduction tolerance ${\rm Tol}$ is then,
$$
n = \left(-\log {\rm Tol}\right) \frac{\sqrt{\kappa}}{2} \sim \left(-\log {\rm Tol}\right) \frac{4}{a m_{ud}},
$$
and thus diverges as the continuum limit is taken at fixed physical quark mass.
  
For Wilson and Wilson-clover Fermions, multigrid algorithms have addressed this critical slowing
down\cite{Luscher:2007se,Brannick:2007ue,Brannick:2007cc,Clark:2008nh,Babich:2009pc,Osborn:2010mb,Frommer:2013kla,Frommer:2013fsa,Frommer:2011bad},
at least for calculation of valence propagators.
The approach is also successful for twisted mass fermions\cite{Alexandrou:2016izb,Alexandrou:2018wiv,Bacchio:2019oiz},
however challenges remain significant for
staggered\cite{Weinberg:2017zlv,Brower:2018ymy,Ayyar:2022krp} and domain wall Fermion/chiral Fermion
discretizations\cite{Cohen:2011ivh,Boyle:2014rwa,Yamaguchi:2016kop,Boyle:2021wcf,Brower:2020xmc}.
Multigrid algorithms will be discussed in more detail in section~\ref{sec:multigrid}.

There remains potential for substantial benefit from reducing the set up cost
of multigrid algorithms, even for the Wilson Fermion discretization,
in order to fully realize the same benefit for gauge configuration generation
\cite{Francis:2019muy,Brower:1995vx}.
An alternative that has been widely used includes the use of both standard eigenvector deflation and also use of
a multigrid compressed local coherence Lanczos for both staggered and domain wall fermions\cite{Clark:2017wom}.
These are algorithmically less exciting and more brute force than multigrid, having a cost of $O(V^2)$ in terms of both
computation and storage, but can be very effective particularly when combined with all-mode-averaging
\cite{Shintani:2014vja,Bali:2009hu}
and the all-to-all volume averaging approach\cite{Foley:2005ac}.
Efficient volume averaging in the low mode space
gives additional statistical advantages \emph{not} necessarily present using multigrid for certain calculations.

\subsection{USQCD SciDAC-5 project}

The USQCD collaboration has recently embarked on a five year funded programme to develop multiscale algorithms
required to exploit the opportunities affored to high energy physics (HEP) by multiple, new exascale computer systems.
The aim is to directly impact the four dimensional QCD HEP simulations being undertaken by USQCD, recognising that these
are currently  predominantly performed using the domain wall and staggered Fermion actions.
Three areas of work were identified: firstly the development of domain wall and staggered Fermion multigrid approaches
with the aim to develop supremacy over existing (multigrid) eigenvector deflation methods.
In this effort the HEP domain scientists are working with leading mathemtical and computer science researchers in the SciDAC FastMath institute.
In particular, Wilson, domain wall and staggered fermion discretisations have been implemented in the PETSc PDE solver package, exposing
lattice gauge theory problems to a broader scientific community.
Secondly the development of
transformational HMC algorithms, displaying some form of successful Fourier acceleration and addressing critical slowing down,
pursuing multiple possible direction including field transformation HMC\cite{Boyle:2022xor},
gauge fixed fourier accelerated HMC\cite{Sheta:2021hsd,PoS:YHuo}
and Riemannian manifold HMC\cite{Cossu:2017eys,PoS:Jung}.
Finally, the development and adoption of domain decomposed HMC evolution for DWF and staggered formulations with the aim of reducing
the sensitivity of our sampling algorithms to computer communication performance\cite{Boyle:2022pai}.
In the long term it will expose our unique problems to a significantly broader
community of algorithmic experts in a programme of joint research.
Some results of the SciDAC effort will be reported in this review, while the idea
that algorithmic research effort accompany the world wide investment in supercomputing
is important to realising the full benefit of the computing.

\section{Acceleration of HMC}

\label{sec:gauge}
\label{sec:FA}

Gauge invariant Fourier acceleration of HMC\cite{Duane:1988vr} was introduced to address critical slowing down,
taking the Fourier analysis point of view appropriate to the ultraviolet limit discussed in section~\ref{sec:CSD}.
Riemannian Manifold HMC\cite{Girolami} (RMHMC) is an independently developed generalisation,
incorporating a generalised field dependent metric function $H[U]$,
weighting the conjugate momentum distribution. The introduction of a symplectic implicit integrator was an important key improvement
over the earlier proposal\cite{Duane:1988vr}.
The HMC integral is re-written as
  $$
  \int d\pi_U \int d\pi_\phi
  \int d \phi
  \int d U \quad
  e^{-\frac{\pi_u H \pi_u}{2}}
  e^{-\frac{\pi_\phi H^{-1} \pi_\phi}{2}}
  e^{-S_{QCD}[U]}
  e^{-\phi^\ast \phi },
  $$
  where the gauge momenta $\pi_U$ are spectrally colored by a functional $H[U]$ and the required Jacobian factor
  is cancelled with an additional, auxiliary Gaussian integral.
  Chulwoo Jung, following earlier work by Guido Cossu has applied RMHMC to four dimensional QCD simulations demonstrating that 
  long distance quantities decorrelate faster with RMHMC, when measured in units Fermion force evaluations\cite{Cossu:2017eys,PoS:Jung}.
  The functional $H[u]$ was originally\cite{Duane:1988vr} a simple low pass filter of the gauge Laplacian operator,
  but has been updated \cite{PoS:Jung} to use arbitrary rational functions that can be shaped to anti-correlate the RMHMC metric
  with the measured force per eigen-mode of the Laplacian and help reduce integration errors.
  However, the implicit integrator on the gauge timestep is considerably more expensive, and a gain in execution time
  has not yet been demonstrated. Further work is required to understand whether this is a software implementation limitation that might
  be addressed with optimisation (or be computer architecture dependent), or if this requires further algorithmic improvement to become
  translate into an execution time gain for QCD.
  
  \begin{figure}[hbt]
    \includegraphics[width=0.5\textwidth]{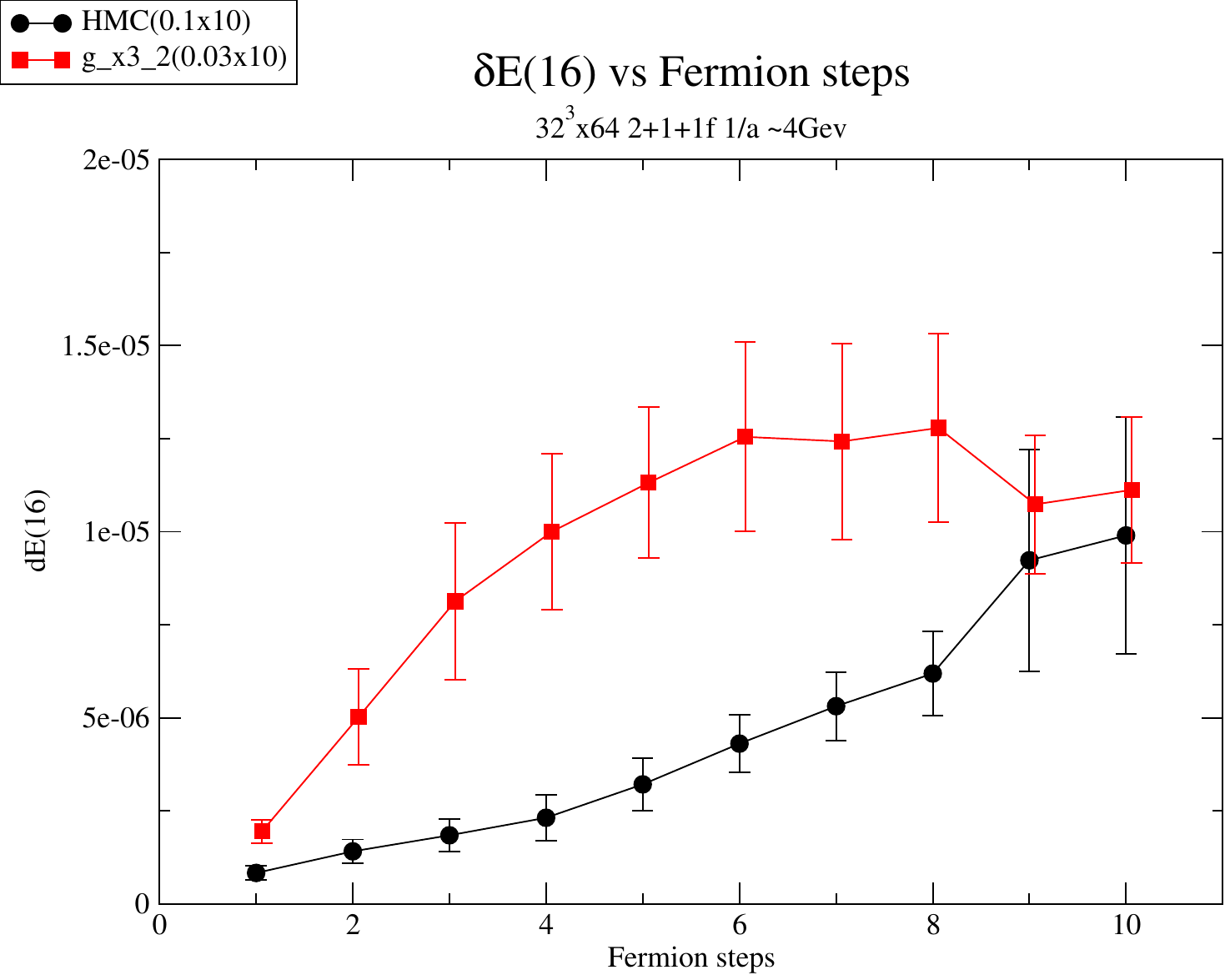}
    \includegraphics[width=0.49\textwidth]{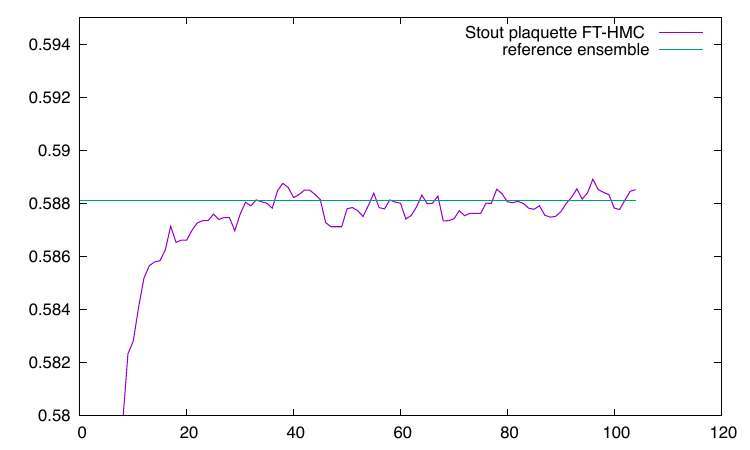}
    \caption{ \label{fig:rmhmc}\label{fig:fthmcplaq}
      Left: Fourier acceleration of HMC can differentially accelerate the decorrelation of infrared distance observables such as the
      Wilson flowed energy $E(16)$ in RMHMC\cite{PoS:Jung}, when measured against the number of Fermion integration timesteps.
      Right: Field transformed HMC evolution with one a single step of stout-like flow in 2+1f DWF evolution, demonstrating
      evolution with the inclusion of Fermions by the USQCD SciDAC-5 project.
    }
  \end{figure}  

  An alternate form of Fourier acceleration involves an approximately fixed (soft covariant) gauge, in
  the gauge fixed fourier accelerated HMC presented by Yikai Huo\cite{PoS:YHuo,Sheta:2021hsd}.
  Since the gauge is (stochastically) fixed, this approach can make use of fast fourier transform for the Fourier acceleration of the
  momentum distribution.
  
\subsection{Flowed HMC algorithms}

There has been a significant growth in activity loosely based on Luscher's Wilson flowed HMC algorithm\cite{Luscher:2010we}.
This loosely can be viewed as being divided into two classes of approach. Complex flows into the deep infra-red, and
even all the way to the trivial strong coupling limit, are being constructed in a way that can
be machine learned producing generative/trivialising maps. Machine learning based lattice
approaches\cite{PoS:Tomiya,PoS:DeSantis,PoS:Sheng,PoS:Gantgen,PoS:Orginos,PoS:Alvestad,PoS:Hackett,PoS:Hoying,PoS:Rossney,PoS:Boyda,PoS:Caselle,PoS:Pederiva,PoS:Finkenrath,PoS:Osborn,PoS:Urban,PoS:Rodekamp,PoS:Abbott,PoS:Foreman,PoS:Wettig,PoS:Wenger,PoS:XYJin,PoS:YLin}
were reviewed separately\cite{Kanwar}.
In this review we consider predominantly {simple flows, targeting the UV region} and retaining a momentum based local
update with field transformation FT-HMC. These directions take the view that
it might be substantially easier to map QCD to QCD at nearby couplings than to trivialise all the way to strong coupling limit.

In this usage, the flowed algorithm can be thought of as a UV smearing function $U(V)$ that
gives {tunable Fourier acceleration} but very much {\emph{incomplete trivialisation}} as a form of change
of variables in the path integral:
    $$
    \int dU e^{-S[ U ]} =     \int dV \left|\frac{dU}{dV} \right| e^{-S[ U(V) ]}.
    $$
    Whereas RMHMC and GFFAHMC change the momentum distribution to ``avoid pushing too hard'' in UV degrees of freedom,
    the FTHMC approach is to retain
    gaussian momenta but introduce a change of variables that differentially ``gears'' UV and IR modes between
    the HMC integration variable $V$ and the physical gauge field $U$.
    In order to be practical, this change of variables must be both cheap and have a cheaply computable Jacobian.
    Transformations targeting the UV region can be more local and so cheaper, but may enable a Fourier acceleration gain.
    While the original Luscher proposal used a continuous Wilson Flow, the SciDAC USQCD project is retaining a discrete
    flow step, similar to Stout smearing, but with $2 N_d$ colored subsets of the links updated one after the other
    to retain a computable Jacobian with stout parameter $\rho=0.1$.

    This was first demonstrated as quenched FT-HMC and using a general set of Wilson loops \cite{Boyle:2022xor}.
    It has been reimplemented for the plaquette flow in Grid\cite{Boyle:2016lbp,Yamaguchi:2022feu}, and is now running in four dimensional QCD with dynamical Fermions
    for the first time. The algorithm is under tuning in the USQCD SciDAC project. An example
    plaquette log is shown reproducing the reference HMC 2+1f plaquette in figure~\ref{fig:fthmcplaq} for $L_s=16$ Domain wall fermions
    with the  Iwasaki gauge action on a $16^3\times 48 $, $\beta=2.13$, $m_{ud}=0.01$, $m_s=0.04$ 2+1 flavour ensemble.
    This made use of TWQCD's exact one flavor algorithm for the strange quark\cite{Chen:2014hyy,Chen:2017gsk}.
    The software was developed under USQCD's SciDAC-5 Workpackage, and was run on AMD GPUs on the ORNL Frontier computer.
    The field transformation overhead is significant but at around 35\%, but sub-dominant compared to the Fermion force.
    The FT-HMC overhead is significantly more scalable than the Fermion solvers
    since the costly elements of the Jacobian force are trivially parallel.

    It remains to be seen whether any (or all) of the Fourier acceleration approaches to HMC (RMHMC, FTHMC, GFFAHMC) lead to a net simulation
  time speed up, however there are clear indications that these methods can in principle address the underlying critical slowing down
  mechanism given the RMHMC results fig~\ref{fig:fthmcplaq}.
  In the case of FTHMC cost overheads are at least close to being small (even at large Fermion mass) making these algorithms
  encouraging.
  
\subsection{Trivialising flows}

  Machine learning sampling algorithms are reviewed elsewhere in this conference\cite{Kanwar}.
  However, some comments here may be appropriate.
  We might expect poor volume and dimensionality scaling for learned proposals, such as machine learned or generative flows,
  if an attempt is made to \emph{fully} trivialize the theory, arising from the volume dependence discussion in section~\ref{sec:dH}.
  A fully trivialized generative algorithm: turning random seeds into gauge configurations
  and obviating the need for costly sampling and data files is perhaps rather optimistic.
  The volume dependence may
  either manifest itself as exponentially poor acceptance rates or small effective sample size, \emph{unless} subvolumes
  of only limited size are being updated.  

  A conventional Markov update makes a local proposal in state space, and while this leads to autocorrelation of configurations,
  the reason for this choice is powerful: having discovered at significant cost the likeliest regions of a very high dimensional
  state space the algorithm naturally tries to explore this region without giving up this knowledge.
  Fortuneately it may be sufficient to drive QCD sampling with a more local flow that resembles UV filtering to address
  critical slowing down, and retain the sequential proposal in Markov state space of HMC.
  
  More complex and learned flows than simple small Wilson loops flow may be advantageous even in such \emph{partial} trivialisation.
  Volume scaling could be addressed if updates/or substitutions on subvolumes of limited size are used to maintain acceptance.
  Since the (pseudo) Fermion action is non-local, effective use of such methods will likely have to mastering some form of
  Fermion determinant domain decomposition as an important element\cite{PoS:Finkenrath}.

\subsection{Parallel tempering and topological tunneling}

\label{sec:PT}

Parallel tempering, also known as replica exchange Monte Carlo, has been successful in addressing critical slowing down
in numerous statistical systems\cite{Swendsen:1986vqb,Marinari:1992qd,Boyd:1997rb,Joo:1998ib}.
The idea is to jointly sample multiple ensembles, including proposals of configuration swapping
between ensembles, with sufficiently nearby parameters (masses, temperatures or couplings) that
a non-zero cross acceptance may be likely.
However early attempted uses in QCD it suffered from poor cross-acceptance probability for physical systems, 
despite attempts at action matching\cite{Joo:1998ib}.

\begin{figure}[hbt]
    \includegraphics[width=0.5\textwidth]{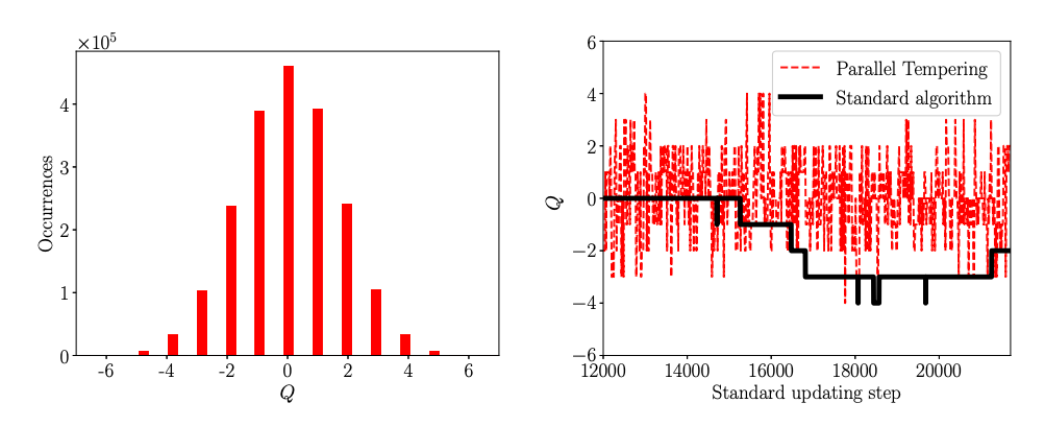}
    \includegraphics[width=0.49\textwidth]{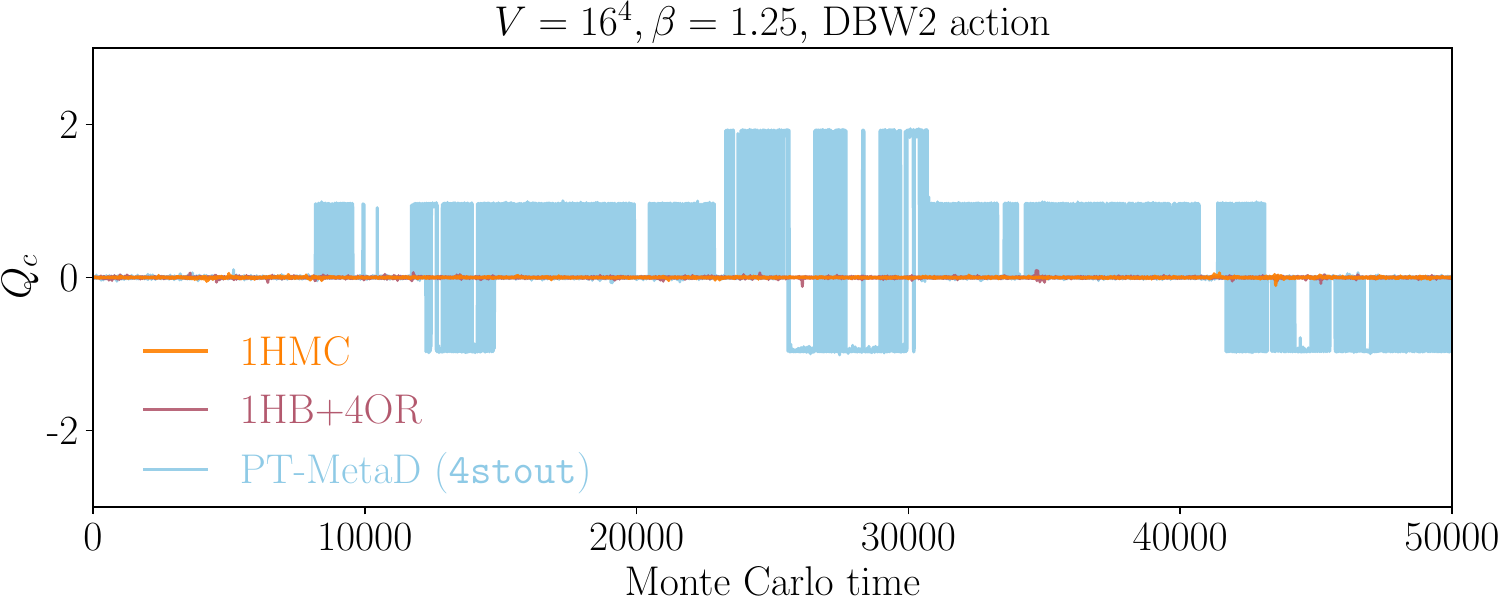}
    \caption{ \label{fig:partemp}
      Left: SU(6) pure gauge evolution with parallel tempered boundary conditions\cite{Bonanno:2022vot}.
      Right: Parallel tempered metadynamics simulation with quenched DBW2 gauge action evolution\cite{Eichhorn:2023uge}.
      In both these algorithms, a topological sampling problem has fixed in quenched 4D simualtion
      \emph{without} changing the physical boundary condition representing the ideal solution to
      the problem (c.f. sec \ref{Sec:ideal}).
    }
  \end{figure}

Since strong coupling simulations to do not develop the problems of ergodicity that result in frozen topology and other
critical slowing down, it is clear that parallel tempering or some other form of communication between strong and weak
coupling simulations can drive topological sampling, and the problem is ``only'' finding an algorithm of sufficient efficiency.
There has been a significant recent effort developing parallel tempered boundary conditions enabling the unconstrained
topological sampling of open boundary conditions in the $CP(N-1)$ model\cite{Hasenbusch:2017unr,Berni:2019bch,Bonanno:2022hmz}
and in quenched $SU(N)$ gauge theory\cite{Bonanno:2020hht,Bonanno:2022vot,Bonanno:2022yjr}, with
impressive improvements to topological sampling
\emph{without} changing they physical boundary condition, fig~\ref{fig:partemp}.
The extension of this to the dynamical fermion case, with the non-local
pseudofermion action is non-trivial but appears to clearly be possible using techniques such as domain decomposition
and it appears these directions may finally provide a \emph{clean} solution to the problem of global topological freezing in the near
future. Results were presented extending this approach to twisted boundary conditions\cite{PoS:DaSilva}.

It was been pointed out in this conference the crossing probability could likely be enhanced by introducing a
(machine learned) field transformation and Jacobian when proposing replica
switching ensembles in parallel tempering\cite{PoS:Hackett}. Further it was proposed
that this could also be performed on subvolumes of the lattice in defect-repair replica exchange (DR-REX).
The idea that subvolumes of size appropriate to a typical physical instanton may be swapped appears intuitively attractive.
The proposals directly addresses concerns about both volume scaling and action parameter differences that have
hampered the historical applications of parallel tempering in QCD simulation.

\begin{figure}[hbt]
\includegraphics[width=0.4\textwidth]{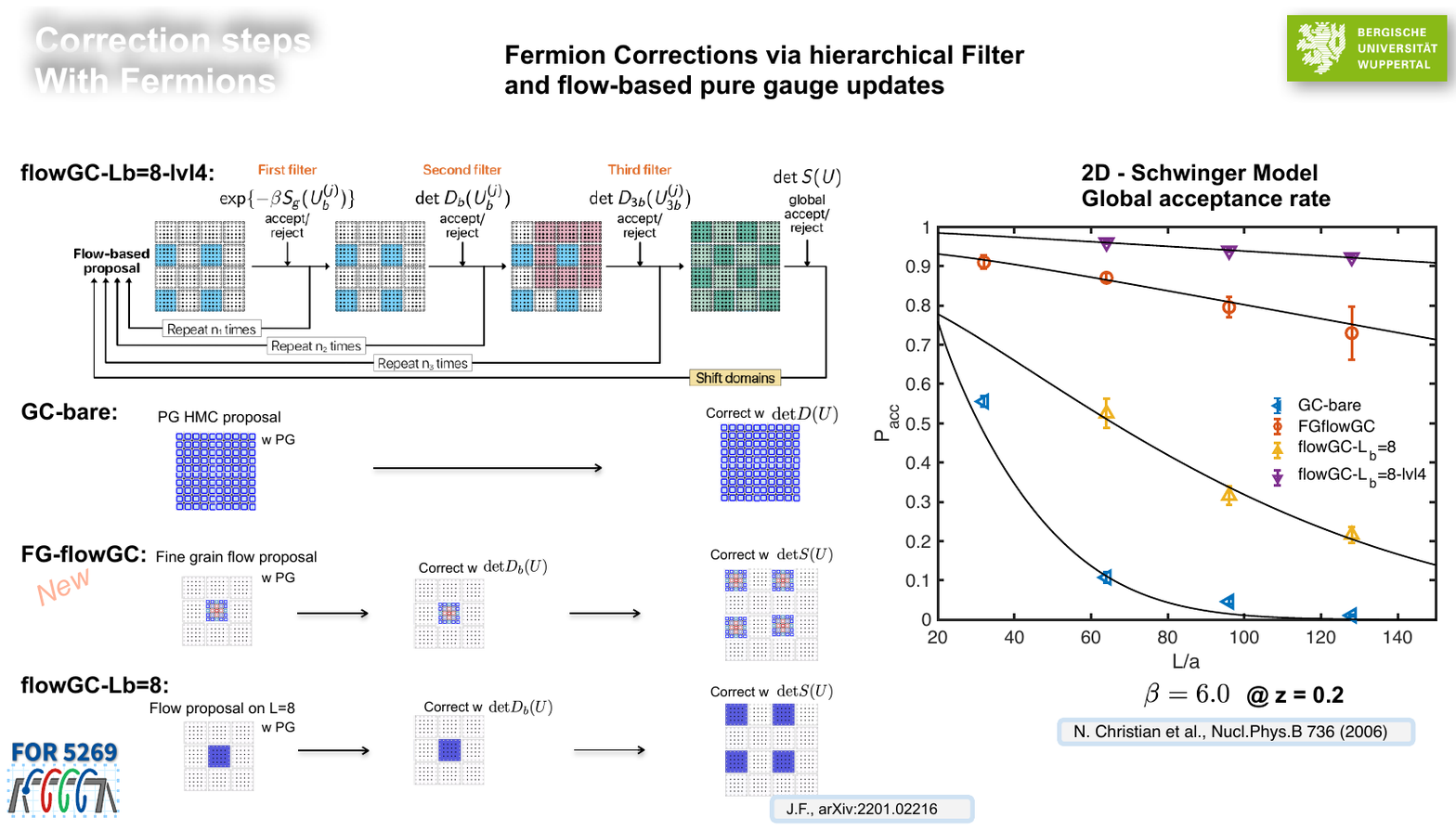}
\includegraphics[width=0.6\textwidth]{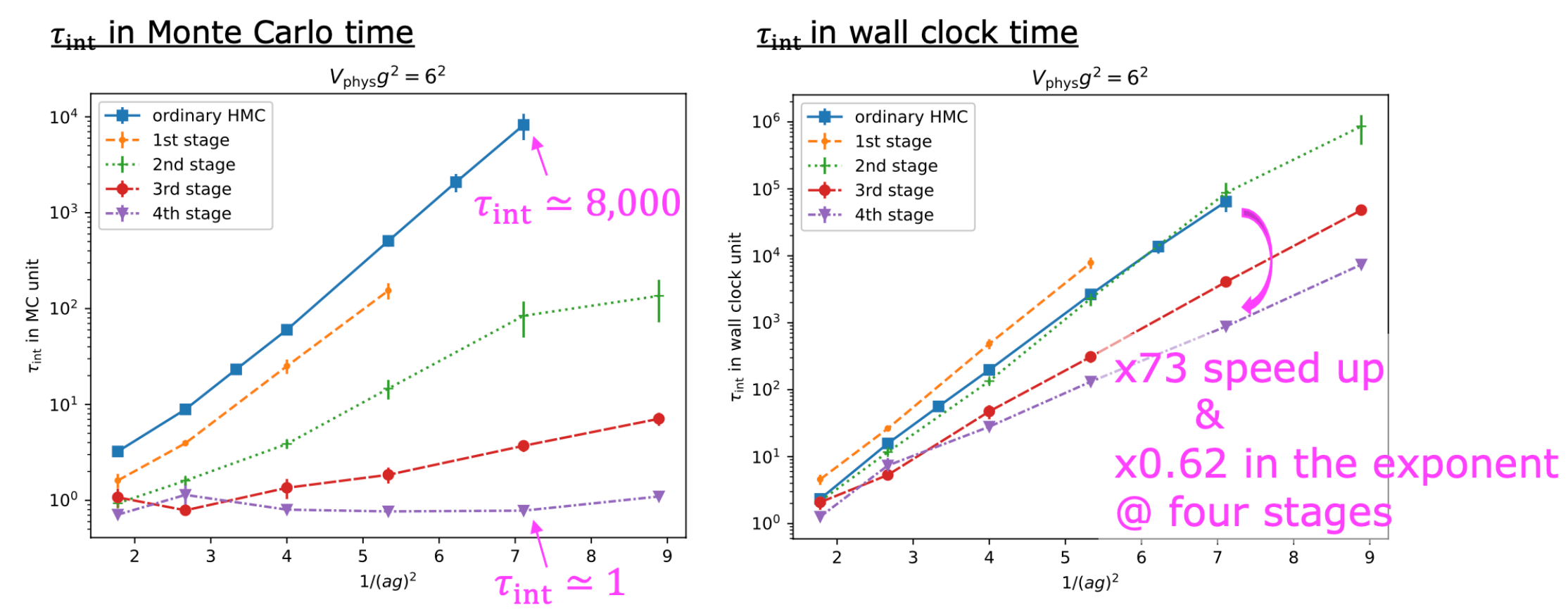}
\caption{
  \label{fig:finkenrath}
  Left: Hierarchical domain decomposed, flowed parallel tempering with fermions in 2D Schwinger model\cite{PoS:Finkenrath}.
  This approach includes many ingredients required for successful treatment of both Fermions \emph{and} good volume scaling,
  and so is very encouraging.
  \label{fig:matsumoto}
  Right: Decimation map accelerates topological sampling in 2D quenched $U(1)$\cite{PoS:Matsumoto}.
}
\end{figure}

Further, work in the 2D Schwinger model addressing the need to include non-local pseudofermion action was presented, using
a well motivated domain decomposition scheme\cite{PoS:Finkenrath} in a heirarchical swapping scheme (figure~\ref{fig:finkenrath}).
This approach seems to include many ingredients required for successful treatment of both Fermions \emph{and} good volume scaling.

Recent methods also connecting to stronger coupling as
a heat bath to drive improved sampling include the use of parallel tempered  metadynamics \cite{Eichhorn:2023uge}, fig~\ref{fig:partemp}, and
out of equilibrium simulation\cite{PoS:Bonnano}.
An approach called the decimation map was introduced in 2D quenched $U(1)$ to integrate out a subset of link
degrees of freedom\cite{PoS:Matsumoto} and produce an effective action, with large acceleration of topological degrees of freedom,
figure~\ref{fig:matsumoto}.

\subsection{Multi-level integration}

A substantial recent effort has been undertaken in the direction of multilevel integration of four dimensional $SU(N)$
gauge theory\cite{Ce:2016idq}, with the inclusion of Fermions\cite{Ce:2016qto} and progression to demonstration
in hadronic correlation functions\cite{Giusti:2018vxm} and
the phenomenologically important and timely calculation of the hadronic vacuum polarization\cite{Giusti:2021qhk}.
The approach (with fermions) uses elements from the domain decomposed HMC.
The lattice is broken into subdomains $\Lambda_0$,  $\Lambda_2$, and an intermediate domain $\Lambda_1$.
A first pass sampling is performed with $N_1$ samples, and these are then held frozen in the
in interstitial domain $\Lambda_1$. For each of these samples, $N_2$ samples are
independently from each of $\Lambda_0$ and $\Lambda_2$ with the other links held fixed, and
a composite set of $N=N_1\times N_2^2 $ gauge configurations may be formed that is quadratic in the cost $N_2$.
The action on each composite configuration can be estimated and reweighted adapting elements of the domain decomposed HMC
to handle the fermion determinant with the reweighting factors bounded by the width of the interstitial domain $\Lambda_1$.

The approach requires the same quadratic enlargement of $N_1\times N_2^2$ valence measurements, and is
most amenable to Wilson-type fermion actions for which multigrid methods are highly efficient, while the approach is
presently somewhat less attractive for domain wall and staggered formulations and calculations using all-to-all methods based
on eigenvector deflation.

 In this conference, multilevel integration results were shown demonstrating exponential
 reduction of signal to noise for the quenched gluebal spectrum \cite{PoS:Barca}, demonstrating the
 potential of these approaches.

\subsection{Master field simulation}

An approach to calculation dubbed master field simulation has recently been
advocated\cite{Luscher:2017cjh,Giusti:2018cmp,Francis:2019muy,Fritzsch:2021klm}.
The proposal observes firstly that in a sufficiently large volume variance of observables is reduced in line
with the volume, an observation known for years as volume averaging, but specifically that when carefully treated the
correlation of observables under translation can be used to infer statistical variance.
Thus \emph{with an equilibrated configuration} in a sufficiently large volume in principle statistical averaging is no-longer
required. This rigorizes heuristic arguments that have been made for a long time about a configuration of macroscopic size.
The second observation is that sensitivity global topology, an issue cutting across many contemporary QCD simulations, is vastly reduced
if $m_\pi L\sim 25$.

Several algorithmic directions have been taken, including a switch from HMC to Stochastic Molecular Dynamics, to address 
integrator stability. The challenge, is of course, equilibrating the lattice in a bounded time and with
sufficient systematic control. There is a certain protection in normal MCMC against non-equilibration because
the variance of the data falls as $\frac{1}{\sqrt{N}}$ while the bias from non-equilibration falls as $\frac{1}{\sqrt{N}}$ in the
sample size. The direction, however is in principle a good one for modern exascale computers which weak scale well but do not strong scale.
It is very likely that the spirit of the approach will manifest itself in future simulations with lower trajectory counts and very much
larger lattice volumes, combined with error estimation making use of the spatial correlation of observables.

\subsection{Domain decomposition}

The DDHMC algorithm has for some time been used, and optimised for CPU architectures
with the subdomains sized to suit a CPU cache memory, and delivered
very high floating point performance for QCD on suitable architectures\cite{Ishikawa:2021iqw}.
The small cell IR regulates Dirichlet solve and limits the cost.
DDHMC has historically been used for two degenerate Wilson light flavours
\cite{Boku:2012zi,PACS-CS:2009sof,PACS-CS:2011ngu,Ishikawa:2015rho,Ishikawa:2021iqw},
The ``boundary'' determinant is projected ($\Pdb$) to exterior boundary of a subdomain,
$$R =  \Pdb - \Pdb \DOi \Dd \DObi \Ddb \quad\quad\quad  R^{-1} = \hPdb - \hPdb D^{-1} \hDdb. $$
This would require a nested solver if used in a rational approximation, and so the strange quark determinant cannot be handled.

A recent proposal has been to use a form of DDHMC with \emph{large} cells, tuned to the largest subvolume that
is natural to process with an \emph{multi-GPU exascale node}\cite{Boyle:2022pai}. This gains from communication avoidance
in systems with large floating point throughput but relatively limited interconnect performance, and also
aligns with the master field idea of substantially larger physical volumes but shorter ensembles.
The infra-red protection of subdomain solver cost is reduced with large cells,
but is partly compensated by a correspondingly reduced force from the boundary determinant.
With such a large domain, broader inactive zones are possible, especially in a master field simulation,
enabling larger force suppression. The perturbative massless zero momentum two point function is proportional
$t^{-3}$ suggesting the force suppression will fall at short distances faster than indicated by a pion mass estimate. 

\begin{figure}[hbt]
  \includegraphics[width=0.48\textwidth]{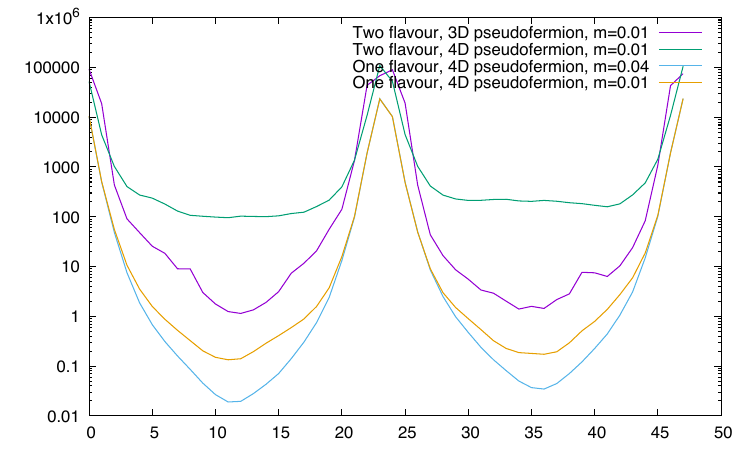}
  \includegraphics[width=0.5\textwidth]{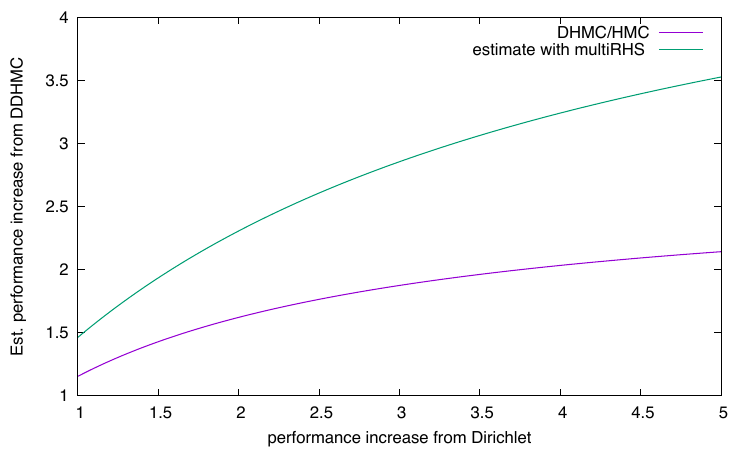}
  \caption{
    \label{fig:ddhmc}
    Left: DDHMC boundary determinant forces within a two domain simulation, with different estimators
    of the determinant. The 4D pseudofermion approach is noiser than the boundary projected pseudofermion, but these forces
    are reduced if a 1+1f approach is taken using the n-roots trick. Domain decomposition with odd $n_F$ is enabled.
    Right:
    Projected speed up is estimated as a function of a ratio of the communicating Dirac operator to the non-communicating Dirac operator.
  }
\end{figure}

The authors remove the boundary projector in the pseudofermion, and rather (writing $\tilde{D}$ as the Dirichlet  operator) use
a determinant ratio four dimensional pseudofermion, rewriting:
$$
\det \left\{ 1 - \DOi \Dd \DObi \Ddb\right\} = \frac{\det D}{\det{\tilde{D}}}.
$$
This introduces a (noiser) four dimensional pseudofermion estimator of the 2 flavour boundary determinant, as
$$
S_{2f} = \phi^\dagger D_{\rm dirichlet} (D^\dagger D)^{-1} D_{\rm dirichlet}^\dagger \phi,
$$
but importantly is in a form where one can now take fractional powers.
Here, the RHMC n-roots trick applied to the boundary determinant reduced forces substantially:
\begin{eqnarray*}
S^B_{1f} &=& \phi_1^\dagger (D_{\rm dirichlet}^\dagger D_{\rm dirichlet})^\frac{1}{4}  
(D^\dagger D)^{-\frac{1}{2}} (D_{\rm dirichlet}^\dagger D_{\rm dirichlet})^\frac{1}{4}  \phi_1,
\end{eqnarray*}
Figure~\ref{fig:ddhmc}, and also enables domain decomposition for odd flavours (i.e. the strange quark).
Using iteration counts from a DWF physical quarm mass evolution, the SciDAC project estimates the speed up as a function of the
ratio of the cost of communicating Dirac operator to the non-communicating Dirac operator, figure~\ref{fig:ddhmc}.

\section{Multigrid and solver algorithms}

\label{sec:multigrid}
Multigrid approaches in Lattice field theory have been transformative for
calculations with Wilson, clover and twisted mass fermion actions.
The key idea is to introduce two complementary preconditioners for the highest and lowest ends of the spectrum.
The low modes are handled in a preconditioner that approximately captures much of the
low mode degrees of freedom and repeatedly reduce these elements in the residual in an outer Krylov
solver. So long as the composite preconditioned system is well conditioned the solver converges rapidly.

The observation that many low modes are locally similar has been made in various ways, namely as the weak approximation
property\cite{Brezina} also called local coherence\cite{Luscher:2007se}.
The basic idea is that in some way a number of vectors are generated that lie within the space spanned by the lowest modes.
These are subsequenty chopped into many disjoint hyper-cuboidal subvectors with some blocking factor.
The span of these sub-block vectors is \emph{vastly} bigger than the span of the original vector set, and we 
empirically find can approximately contain the complete set of lowest modes due to the observation of local coherence.
Since eigenvalues are gauge invariant but eigenvectors are covariant, the obvious
way to generate these vectors and so ``define'' a local averaging in a block is by
using the covariant Dirac operator itself to filter out
the near null space of the operator.
``Important'' basis vectors $\phi_k(x)$ lying in the near null space are
typically generated by repeated application of an approximate inverse of
the matrix to a random vector, perhaps using the first multigrid solver to improve it's own subspace, and more
generally can be thought of a spectral filtering problem.
Once obtained, these vectors are obtained, they are restricted to disjoint blocks $b$ and orthonormalised. The
span of these block vectors  $\phi_k^b(x)$ is much larger than the span of the $\phi_k(x)$, and due to local coherence
includes many low eigenmodes as a subspace. The sharp edges however introduce high mode spectral leakage when a
coarse grid correction is applied and the composite preconditioner performs best if fine-grid post-smoothing is included.
Coarse degrees of freedom defined by inner product with defining a projector to this subspace $S$, and it's complement $\bar{S}$.
\beq
P_S =  \sum_{k,b} |\phi^b_k\rangle \langle \phi^b_k | \quad\quad ; \quad\quad P_{\bar{S}} = 1 - P_S 
\eeq
A faithful representation of the Dirac matrix may beformed within the subspace spanned by these $|\phi_k^b\rangle$
by calculating all non-zero matrix elements of the operator.
\beq
M=
\left(
\begin{array}{cc}
M_{\bar{S}\bar{S}} & M_{S\bar{S}}\\
M_{\bar{S}S} &M_{SS}
\end{array}
\right)=
\left(
\begin{array}{cc}
P_{\bar{S}} M P_{\bar{S}}  &  P_S M P_{\bar{S}}\\
 P_{\bar{S}} M P_S &   P_S M P_S
\end{array}
\right)
\eeq
The coarse grid matrix is also known as the \emph{little Dirac operator},
\beq
A^{ab}_{jk} = \langle \phi^a_j| M | \phi^b_k\rangle
\quad\quad ; \quad\quad
(M_{SS}) = A_{ij}^{ab} |\phi_i^a\rangle \langle \phi_j^b |.
\eeq
and the subspace inverse can be solved by Krylov methods.
It is important to note that $A$ inherits a sparse structure from $M$ because well separated blocks do \emph{not} connect through $M$.

In the case of Wilson, clover and twisted mass Fermions the Dirac operator is one-hop and spectrally amenable to
invert with non-Hermitian Krylov solver algorithms, such as GCR and BiCGstab.
The five dimensional matrix of Mobius and Domain wall Fermions have a spectrum that completely encloses the origin in the complex
plain due to the large negative Wilson mass term. In the infinite volume limit, this spectrum is dense and a Krylov solver
must use a polynomial to approximate $1/z$ over this domain of the spectrum. Thus\cite{Boyle:2021wcf} there is an integral one can perform
on any possible Krylov polynomial that differs from the true solution so that these must differ. This manifests itself in very slow
convergence (of order the system size). The observation is commonly stated as non-hermitian Krylov solvers having
a ``half plane condition''\cite{Trefethen} such that all eigenvalues should lie in one half of the complex plane.

The consequence of this is that, to date, only four approaches to Domain Wall Fermions have had any measure of success:
squaring the operator and coarsening a 2-hop (unpreconditioned squared matrix)\cite{Cohen:2011ivh};
or 4-hop (red-black preconditioned squared operator)\cite{Boyle:2014rwa}; and
restricting the action to the $c=0$ subspace of Mobius and coarsening the hermitian indefinite operator (with real spectrum)
$\Gamma_5 R_5 D_{dwf}$\cite{Yamaguchi:2016kop,Boyle:2021wcf};
and finally coarsening both the Pauli Villars \emph{and} the Domain wall operator and solving the
combination $D(m=1)^\dagger D(m=m_{ud})$ in a non-Hermitian preconditioned system that successfully transforms the
spectrum to a half plane\cite{Brower:2020xmc}.

HISQ fermions include a three hop Naik term. Fortunately, if the cell size is at least $3^4$ this does not imply more than one hop in the coarse space.
Significant progress has been made with Kahler Dirac preconditioning \cite{Brower:2018ymy,Ayyar:2022krp}, however further research continues under the SciDAC-5 project to make
a multigrid that becomes the method of choice for staggered valence analysis compared to (say) eigenvector deflation.
An additive Schwartz preconditioner was introduced for staggered Fermions at this conference\cite{PoS:Weinberg}.

Under SciDAC-5 the Grid software package\cite{Boyle:2016lbp,Yamaguchi:2022feu}
has been updated to support 1-hop, 2-hop and 4-hop coarsening schemes connecting
a general coarse matrix stencil ranging from displacements $(-1,-1,-1,-1)$ to $(1,1,1,1)$ for up to $3^3=81$ points in the stencil.
This is performed using a ``ghost zone'' halo exchange of depth one, and the communication cost does not particularly grow with the
stencil size. When the number of basis vectors is large the volume of data in matrix elements $A_{ij}$ is much greater than the volume
of data in a coarse vector and the cost of applying the coarse operator is dominated by the cost of fetching these coefficient matrices.
The coarse operator implementation in Grid has therefore been updated to support application
to multiple right hand sides concurrently. Further, there are relatively new, convenient and well optimised
machine learning targeted ``batched BLAS GEMM'' routines
for performing a list of many general matrix-matrix multiplications.
Software interfaces are available under CUDA, HIP, and SYCL interfaces easing portability across
modern GPUs. The arithmetic for a single displacement in the stencil of the coarse operator can be posed as a batched GEMM call,
$$
C_{N_{basis} \times N_{rhs}}(x)
= 
C_{N_{basis} \times N_{rhs}}(x) +
A^p_{N_{basis} \times N_{basis}}(x) \times
B_{N_{basis} \times N_{rhs}}(x+\delta_p),
$$
and it is relatively easy to obtain O(24) Tflop/s per node in double precision on the Frontier supercomputer (for example).
By using these interfaces, access is given to tensor multiplication hardware with
significantly increased peak and actual floating point throughput.
The performance is good because the matrix ranks are significantly larger than arises with $N_c = 3$ fine grid operations.
It is wise for Lattice software to take the opportunity to use machine learning targeted hardware when appropriate.

The SciDAC project's multiple right hand side version of HDCG\cite{Boyle:2014rwa} currently gives a factor of around five fold
gain over on $48^3$ at physical quark masses with domain wall fermions.
This is projected to rise to nearer fifteen fold when a coarse grid deflation bottleneck
is also implemented using batched GEMM, and the algorithm is intended
to accelerate the calculation of the hadronic vacuum polarisation
on volumes that are too large for eigenvector deflation to be tractable.
Flexible stencil patterns (9 point, 33 point and 81 point) are supported to enable coarsening of a number of Fermion
operators (and products of them\cite{Brower:2020xmc}).
Fast subspace setup using chebyshev polynomial filters\cite{Yamaguchi:2016kop} is being investigated.

\begin{figure}[hbt]
\includegraphics[width=0.68\textwidth]{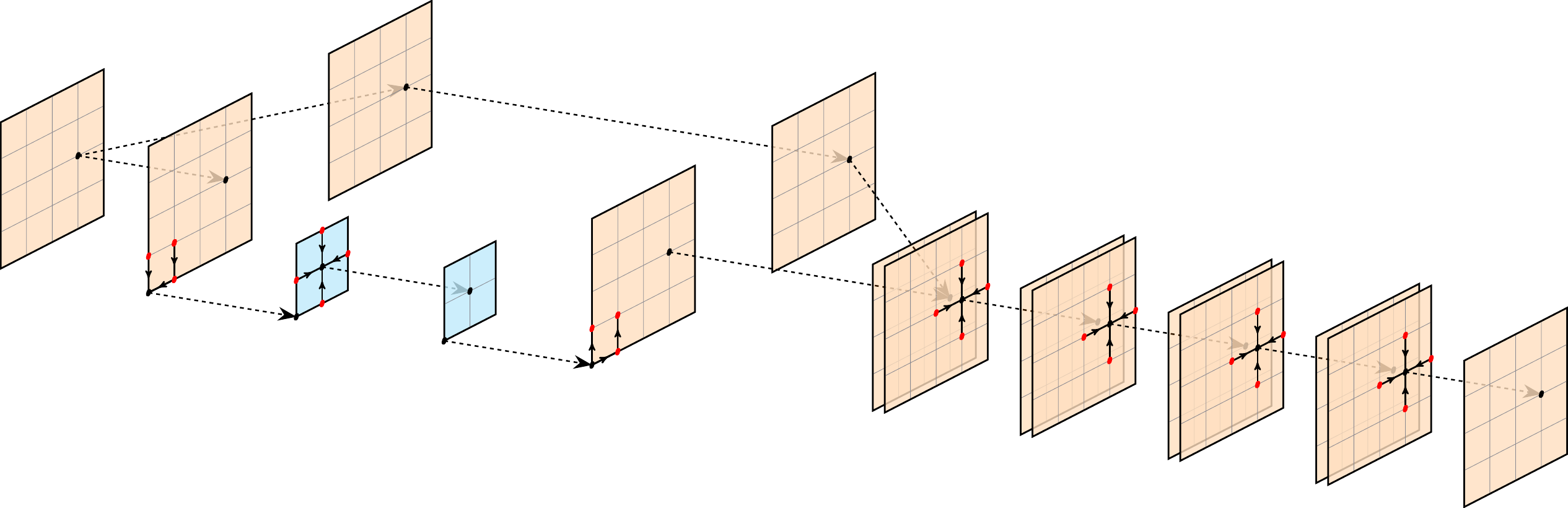}
\includegraphics[width=0.3\textwidth]{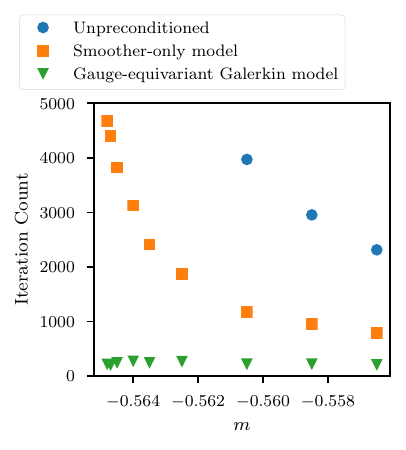}
\caption{
  \label{fig:equivariant}
  Left: Gauge equivariant neural networks are trained\cite{PoS:Wettig} with parallel transport stencil layers to directly
  approximate the inverse of the Dirac operator on a coarse grid. This differs from conventional lattice multigrid where
  the operator itself is approximated and an iterative inversion of the approximation is used in the coarse grid correction.
  Right: Elimination of Fermionic critical slowing down on a Wilson-clover test system\cite{PoS:Wettig}.
}
\end{figure}

A neural network based multigrid was presented \cite{PoS:Wettig,Lehner:2023bba,Lehner:2023prf}.
This uses a covariant (called equivariant) parallel transport layer to build a large stencil
operator in a neural network, figure~\ref{fig:equivariant}. The space of multilayer networks contains a Krylov polynomial
of a coarse grid operator as subspace, so this can at some level be view as a generalisation or
superset of conventional lattice coarse grid corrections. However, from another perspective
their neural network idea replaces an iterative Krylov or polynomial of a coarse grid operator
designed to represent the Dirac operator, with a \emph{larger} coarse grid object trained to directly
approximate the \emph{inverse} of the Dirac operator. On small volumes the method works well from
an algorithmic perspective, with both covariant high and low mode preconditioners demonstrating efficacy and
also reduced cost transfer of set up cost across gauge configurations.
It will be interesting to see if the approach scales to larger volumes and light quark masses with a
gain in time to solution.

\section{Conclusions}

The author is pleased to comment that there are a significant number of promising directions being pursued by the field.
Firstly, the recent development of (quenched) parallel tempered boundary
conditions\cite{Hasenbusch:2017unr,Berni:2019bch,Bonanno:2022hmz,Bonanno:2020hht,Bonanno:2022vot,Bonanno:2022yjr}
(supplemented by related defect-repair replica exchange\cite{PoS:Hackett}
or local updates \cite{PoS:Finkenrath}) offer a path to resolving global topological freezing.
Possibly combined with some form of Fourier acceleration or flowed HMC there are routes that may plausibly
address critical slowing down in gauge evolution. Clearly continued investment algorithmic research
is required and very much worthwhile. 
The current focus of the field on trivialisation may be more ambitious than is required to solve our problems.
It is surely easier to ``match'' the distribution of gauge fields in an evolution at the weakest coupling for which we ergodically sample
to an even weaker coupling than to trivialise all the way to the strong coupling limit. The locality and cost of the corresponding flow
is likely favourable and the author does not believe that the volume scaling of effective sample sizes will make entirely dropping HMC based
evolution a likely outcome. New multigid approaches are being studied for domain wall and staggered fermions, while new
approaches to use covariant neural networks to directly approximate the coarse space inverse, and avoid , have been proposed.
In multigrid, it has been established that multiple right-hand-side solvers can reduce the cost of the course space in multigrid
substantially and align the arithmetic operations in a large part of QCD
calculations with the hardware that has emerged for machine learning, delivering very high performance.

\section{Acknowledgements}

I wish to thank my colleagues in RBC-UKQCD and in the USQCD SciDAC project for many useful conversations.
PB has been supported in part by the Scientific Discovery through
Advanced Computing (SciDAC) program LAB 22-2580, and by US
DOE Contract DESC0012704(BNL).


\begin{thebibliography}{99}

\bibitem{Kanwar}
  G. Kanwar
  ``Flow-based sampling for lattice field theories''
  PoS(LATTICE2023)114 

\bibitem{Metropolis:1953am}
N.~Metropolis, A.~W.~Rosenbluth, M.~N.~Rosenbluth, A.~H.~Teller and E.~Teller,
``Equation of state calculations by fast computing machines,''
J. Chem. Phys. \textbf{21} (1953), 1087-1092
doi:10.1063/1.1699114

\bibitem{Duane:1987de}
S.~Duane, A.~D.~Kennedy, B.~J.~Pendleton and D.~Roweth,
``Hybrid Monte Carlo,''
Phys. Lett. B \textbf{195} (1987), 216-222
doi:10.1016/0370-2693(87)91197-X



\bibitem{Horowitz:1985kd}
A.~M.~Horowitz,
``Stochastic Quantization in Phase Space,''
Phys. Lett. B \textbf{156} (1985), 89
doi:10.1016/0370-2693(85)91360-7



\bibitem{Clark:2004cp}
M.~A.~Clark, A.~D.~Kennedy and Z.~Sroczynski,
``Exact 2+1 flavour RHMC simulations,''
Nucl. Phys. B Proc. Suppl. \textbf{140} (2005), 835-837
doi:10.1016/j.nuclphysbps.2004.11.192
[arXiv:hep-lat/0409133 [hep-lat]].

\bibitem{Clark:2003na}
M.~A.~Clark and A.~D.~Kennedy,
``The RHMC algorithm for two flavors of dynamical staggered fermions,''
Nucl. Phys. B Proc. Suppl. \textbf{129} (2004), 850-852
doi:10.1016/S0920-5632(03)02732-4
[arXiv:hep-lat/0309084 [hep-lat]].


\bibitem{Clark:2006wq}
M.~A.~Clark,
``The Rational Hybrid Monte Carlo Algorithm,''
PoS \textbf{LAT2006} (2006), 004
doi:10.22323/1.032.0004
[arXiv:hep-lat/0610048 [hep-lat]].


\bibitem{Mohler:2020txx}
D.~Mohler and S.~Schaefer,
``Remarks on strange-quark simulations with Wilson fermions,''
Phys. Rev. D \textbf{102} (2020) no.7, 074506
doi:10.1103/PhysRevD.102.074506
[arXiv:2003.13359 [hep-lat]].



\bibitem{Chen:2014hyy}
Y.~C.~Chen \textit{et al.} [TWQCD],
``Exact Pseudofermion Action for Monte Carlo Simulation of Domain-Wall Fermion,''
Phys. Lett. B \textbf{738} (2014), 55-60
doi:10.1016/j.physletb.2014.09.016
[arXiv:1403.1683 [hep-lat]].


\bibitem{Chen:2017gsk}
Y.~C.~Chen and T.~W.~Chiu,
``Mass Preconditioning for the Exact One-Flavor Action in Lattice QCD with Domain-Wall Fermion,''
[arXiv:1710.09621 [hep-lat]].



\bibitem{Clark:2005sq}
M.~A.~Clark, P.~de Forcrand and A.~D.~Kennedy,
``Algorithm shootout: R versus RHMC,''
PoS \textbf{LAT2005} (2006), 115
doi:10.22323/1.020.0115
[arXiv:hep-lat/0510004 [hep-lat]].

\bibitem{Gottlieb:1987mq}
S.~A.~Gottlieb, W.~Liu, D.~Toussaint, R.~L.~Renken and R.~L.~Sugar,
``Hybrid Molecular Dynamics Algorithms for the Numerical Simulation of Quantum Chromodynamics,''
Phys. Rev. D \textbf{35} (1987), 2531-2542
doi:10.1103/PhysRevD.35.2531

\bibitem{Francis:2019muy}
A.~Francis, P.~Fritzsch, M.~L\"uscher and A.~Rago,
``Master-field simulations of O($a$)-improved lattice QCD: Algorithms, stability and exactness,''
Comput. Phys. Commun. \textbf{255} (2020), 107355
doi:10.1016/j.cpc.2020.107355
[arXiv:1911.04533 [hep-lat]].

\bibitem{Hasenbusch:2001ne}
M.~Hasenbusch,
``Speeding up the hybrid Monte Carlo algorithm for dynamical fermions,''
Phys. Lett. B \textbf{519} (2001), 177-182
doi:10.1016/S0370-2693(01)01102-9
[arXiv:hep-lat/0107019 [hep-lat]].

\bibitem{Urbach:2005ji}
C.~Urbach, K.~Jansen, A.~Shindler and U.~Wenger,
``HMC algorithm with multiple time scale integration and mass preconditioning,''
Comput. Phys. Commun. \textbf{174} (2006), 87-98
doi:10.1016/j.cpc.2005.08.006
[arXiv:hep-lat/0506011 [hep-lat]].


\bibitem{Luscher:2004pav}
M.~Luscher,
``Schwarz-preconditioned HMC algorithm for two-flavour lattice QCD,''
Comput. Phys. Commun. \textbf{165} (2005), 199-220
doi:10.1016/j.cpc.2004.10.004
[arXiv:hep-lat/0409106 [hep-lat]].

\bibitem{DelDebbio:2006cn}
L.~Del Debbio, L.~Giusti, M.~Luscher, R.~Petronzio and N.~Tantalo,
``QCD with light Wilson quarks on fine lattices (I): First experiences and physics results,''
JHEP \textbf{02} (2007), 056
doi:10.1088/1126-6708/2007/02/056
[arXiv:hep-lat/0610059 [hep-lat]].


\bibitem{DelDebbio:2007pz}
L.~Del Debbio, L.~Giusti, M.~Luscher, R.~Petronzio and N.~Tantalo,
``QCD with light Wilson quarks on fine lattices. II. DD-HMC simulations and data analysis,''
JHEP \textbf{02} (2007), 082
doi:10.1088/1126-6708/2007/02/082
[arXiv:hep-lat/0701009 [hep-lat]].



\bibitem{Clark:2006fx}
M.~A.~Clark and A.~D.~Kennedy,
``Accelerating dynamical fermion computations using the rational hybrid Monte Carlo (RHMC) algorithm with multiple pseudofermion fields,''
Phys. Rev. Lett. \textbf{98} (2007), 051601
doi:10.1103/PhysRevLett.98.051601
[arXiv:hep-lat/0608015 [hep-lat]].

\bibitem{deForcrand:2018orx}
P.~de Forcrand and L.~Keegan,
``Rational hybrid Monte Carlo with block solvers and multiple pseudofermions,''
Phys. Rev. E \textbf{98} (2018) no.4, 043306
doi:10.1103/PhysRevE.98.043306
[arXiv:1808.01829 [hep-lat]].



\bibitem{Kennedy:2000ju}
A.~D.~Kennedy and B.~Pendleton,
``Cost of the generalized hybrid Monte Carlo algorithm for free field theory,''
Nucl. Phys. B \textbf{607} (2001), 456-510
doi:10.1016/S0550-3213(01)00129-8
[arXiv:hep-lat/0008020 [hep-lat]].

\bibitem{Kennedy:1999di}
A.~D.~Kennedy and B.~Pendleton,
``Cost of generalized HMC algorithms for free field theory,''
Nucl. Phys. B Proc. Suppl. \textbf{83} (2000), 816-818
doi:10.1016/S0920-5632(00)91813-9
[arXiv:hep-lat/0001031 [hep-lat]].

\bibitem{Sheta:2021hsd}
A.~Sheta, Y.~Zhao and N.~H.~Christ,
``Gauge-Fixed Fourier Acceleration,''
PoS \textbf{LATTICE2021} (2022), 084
doi:10.22323/1.396.0084
[arXiv:2108.05486 [hep-lat]].




\bibitem{DelDebbio:2002xa}
L.~Del Debbio, H.~Panagopoulos and E.~Vicari,
``theta dependence of SU(N) gauge theories,''
JHEP \textbf{08} (2002), 044
doi:10.1088/1126-6708/2002/08/044
[arXiv:hep-th/0204125 [hep-th]].

\bibitem{DelDebbio:2004xh}
L.~Del Debbio, G.~M.~Manca and E.~Vicari,
``Critical slowing down of topological modes,''
Phys. Lett. B \textbf{594} (2004), 315-323
doi:10.1016/j.physletb.2004.05.038
[arXiv:hep-lat/0403001 [hep-lat]].



\bibitem{Luscher:2010we}
M.~Luscher,
``Topology, the Wilson flow and the HMC algorithm,''
PoS \textbf{LATTICE2010} (2010), 015
doi:10.22323/1.105.0015
[arXiv:1009.5877 [hep-lat]].

\bibitem{Schaefer:2010hu}
S.~Schaefer \textit{et al.} [ALPHA],
``Critical slowing down and error analysis in lattice QCD simulations,''
Nucl. Phys. B \textbf{845} (2011), 93-119
doi:10.1016/j.nuclphysb.2010.11.020
[arXiv:1009.5228 [hep-lat]].

\bibitem{Luscher:2011kk}
M.~Luscher and S.~Schaefer,
``Lattice QCD without topology barriers,''
JHEP \textbf{07} (2011), 036
doi:10.1007/JHEP07(2011)036
[arXiv:1105.4749 [hep-lat]].

\bibitem{McGlynn:2014bxa}
G.~McGlynn and R.~D.~Mawhinney,
``Diffusion of topological charge in lattice QCD simulations,''
Phys. Rev. D \textbf{90} (2014) no.7, 074502
doi:10.1103/PhysRevD.90.074502
[arXiv:1406.4551 [hep-lat]].




\bibitem{saad}
Saad, Y. (2003). Iterative Methods for Sparse Linear Systems. SIAM. ISBN: 978-0-89871-534-7

\bibitem{Luscher:2007se}
M.~Luscher,
``Local coherence and deflation of the low quark modes in lattice QCD,''
JHEP \textbf{07} (2007), 081
doi:10.1088/1126-6708/2007/07/081
[arXiv:0706.2298 [hep-lat]].

\bibitem{Brannick:2007ue}
J.~Brannick, R.~C.~Brower, M.~A.~Clark, J.~C.~Osborn and C.~Rebbi,
``Adaptive Multigrid Algorithm for Lattice QCD,''
Phys. Rev. Lett. \textbf{100} (2008), 041601
doi:10.1103/PhysRevLett.100.041601
[arXiv:0707.4018 [hep-lat]].

\bibitem{Brannick:2007cc}
J.~Brannick, R.~C.~Brower, M.~A.~Clark, J.~C.~Osborn and C.~Rebbi,
``Adaptive Multigrid Algorithm for the QCD Dirac-Wilson Operator,''
PoS \textbf{LATTICE2007} (2007), 029
doi:10.22323/1.042.0029
[arXiv:0710.3612 [hep-lat]].

\bibitem{Clark:2008nh}
M.~A.~Clark, J.~Brannick, R.~C.~Brower, S.~F.~McCormick, T.~A.~Manteuffel, J.~C.~Osborn and C.~Rebbi,
``The Removal of critical slowing down,''
PoS \textbf{LATTICE2008} (2008), 035
doi:10.22323/1.066.0035
[arXiv:0811.4331 [hep-lat]].

\bibitem{Babich:2009pc}
R.~Babich, J.~Brannick, R.~C.~Brower, M.~A.~Clark, S.~D.~Cohen, J.~C.~Osborn and C.~Rebbi,
``The Role of multigrid algorithms for LQCD,''
PoS \textbf{LAT2009} (2009), 031
doi:10.22323/1.091.0031
[arXiv:0912.2186 [hep-lat]].

\bibitem{Osborn:2010mb}
J.~C.~Osborn, R.~Babich, J.~Brannick, R.~C.~Brower, M.~A.~Clark, S.~D.~Cohen and C.~Rebbi,
``Multigrid solver for clover fermions,''
PoS \textbf{LATTICE2010} (2010), 037
doi:10.22323/1.105.0037
[arXiv:1011.2775 [hep-lat]].

\bibitem{Frommer:2013kla}
  A.~Frommer, K.~Kahl, S.~Krieg, B.~Leder and M.~Rottmann,
  ``An adaptive aggregation based domain decomposition multilevel method for the lattice wilson dirac operator: multilevel results,''
  [arXiv:1307.6101 [hep-lat]].
\bibitem{Frommer:2013fsa}
A.~Frommer, K.~Kahl, S.~Krieg, B.~Leder and M.~Rottmann,
``Adaptive Aggregation-Based Domain Decomposition Multigrid for the Lattice Wilson--Dirac Operator,''
SIAM J. Sci. Comput. \textbf{36} (2014) no.4, A1581-A1608
doi:10.1137/130919507
[arXiv:1303.1377 [hep-lat]].
\bibitem{Frommer:2011bad}
A.~Frommer, K.~Kahl, S.~Krieg, B.~Leder and M.~Rottmann,
``Aggregation-based Multilevel Methods for Lattice QCD,''
PoS \textbf{LATTICE2011} (2011), 046
doi:10.22323/1.139.0046
[arXiv:1202.2462 [hep-lat]].


\bibitem{Alexandrou:2016izb}
C.~Alexandrou, S.~Bacchio, J.~Finkenrath, A.~Frommer, K.~Kahl and M.~Rottmann,
``Adaptive Aggregation-based Domain Decomposition Multigrid for Twisted Mass Fermions,''
Phys. Rev. D \textbf{94} (2016) no.11, 114509
doi:10.1103/PhysRevD.94.114509
[arXiv:1610.02370 [hep-lat]].

\bibitem{Bacchio:2019oiz}
S.~G.~Bacchio,
``Simulating maximally twisted fermions at the physical pointa with multigrid methods,''
doi:10.25926/jy14-2238

\bibitem{Alexandrou:2018wiv}
C.~Alexandrou, S.~Bacchio and J.~Finkenrath,
``Multigrid approach in shifted linear systems for the non-degenerated twisted mass operator,''
Comput. Phys. Commun. \textbf{236} (2019), 51-64
doi:10.1016/j.cpc.2018.10.013
[arXiv:1805.09584 [hep-lat]].

\bibitem{Weinberg:2017zlv}
E.~S.~Weinberg, R.~C.~Brower, K.~Clark and A.~Strelchenko,
``Progress Report on Staggered Multigrid,''
PoS \textbf{LATTICE2016} (2017), 273
doi:10.22323/1.256.0273
\bibitem{Brower:2018ymy}
R.~C.~Brower, M.~A.~Clark, A.~Strelchenko and E.~Weinberg,
``Multigrid algorithm for staggered lattice fermions,''
Phys. Rev. D \textbf{97} (2018) no.11, 114513
doi:10.1103/PhysRevD.97.114513
[arXiv:1801.07823 [hep-lat]].
\bibitem{Ayyar:2022krp}
V.~Ayyar, R.~C.~Brower, M.~A.~Clark, M.~Wagner and E.~Weinberg,
``Optimizing Staggered Multigrid for Exascale performance,''
PoS \textbf{LATTICE2022} (2023), 335
doi:10.22323/1.430.0335
[arXiv:2212.12559 [hep-lat]].


\bibitem{Cohen:2011ivh}
S.~D.~Cohen, R.~C.~Brower, M.~A.~Clark and J.~C.~Osborn,
``Multigrid Algorithms for Domain-Wall Fermions,''
PoS \textbf{LATTICE2011} (2011), 030
doi:10.22323/1.139.0030
[arXiv:1205.2933 [hep-lat]].

\bibitem{Boyle:2014rwa}
P.~A.~Boyle,
[arXiv:1402.2585 [hep-lat]].

\bibitem{Yamaguchi:2016kop}
A.~Yamaguchi and P.~Boyle,
``Hierarchically deflated conjugate residual,''
PoS \textbf{LATTICE2016} (2016), 374
doi:10.22323/1.256.0374
[arXiv:1611.06944 [hep-lat]].

\bibitem{Brower:2020xmc}
R.~C.~Brower, M.~A.~Clark, D.~Howarth and E.~S.~Weinberg,
``Multigrid for chiral lattice fermions: Domain wall,''
Phys. Rev. D \textbf{102} (2020) no.9, 094517
doi:10.1103/PhysRevD.102.094517
[arXiv:2004.07732 [hep-lat]].

\bibitem{Boyle:2021wcf}
P.~Boyle and A.~Yamaguchi,
``Comparison of Domain Wall Fermion Multigrid Methods,''
[arXiv:2103.05034 [hep-lat]].


\bibitem{Brower:1995vx}
R.~C.~Brower, T.~Ivanenko, A.~R.~Levi and K.~N.~Orginos,
``Chronological inversion method for the Dirac matrix in hybrid Monte Carlo,''
Nucl. Phys. B \textbf{484} (1997), 353-374
doi:10.1016/S0550-3213(96)00579-2
[arXiv:hep-lat/9509012 [hep-lat]].



\bibitem{Clark:2017wom}
M.~A.~Clark, C.~Jung and C.~Lehner,
``Multi-Grid Lanczos,''
EPJ Web Conf. \textbf{175} (2018), 14023
doi:10.1051/epjconf/201817514023
[arXiv:1710.06884 [hep-lat]].

\bibitem{Shintani:2014vja}
E.~Shintani, R.~Arthur, T.~Blum, T.~Izubuchi, C.~Jung and C.~Lehner,
``Covariant approximation averaging,''
Phys. Rev. D \textbf{91} (2015) no.11, 114511
doi:10.1103/PhysRevD.91.114511
[arXiv:1402.0244 [hep-lat]].

\bibitem{Bali:2009hu}
G.~S.~Bali, S.~Collins and A.~Schafer,
``Effective noise reduction techniques for disconnected loops in Lattice QCD,''
Comput. Phys. Commun. \textbf{181} (2010), 1570-1583
doi:10.1016/j.cpc.2010.05.008
[arXiv:0910.3970 [hep-lat]].

\bibitem{Foley:2005ac}
J.~Foley, K.~Jimmy Juge, A.~O'Cais, M.~Peardon, S.~M.~Ryan and J.~I.~Skullerud,
``Practical all-to-all propagators for lattice QCD,''
Comput. Phys. Commun. \textbf{172} (2005), 145-162
doi:10.1016/j.cpc.2005.06.008
[arXiv:hep-lat/0505023 [hep-lat]].
  


\bibitem{Boyle:2022xor}
P.~Boyle, T.~Izubuchi, L.~Jin, C.~Jung, C.~Lehner, N.~Matsumoto and A.~Tomiya,
``Use of Schwinger-Dyson equation in constructing an approximate trivializing map,''
PoS \textbf{LATTICE2022} (2023), 229
doi:10.22323/1.430.0229
[arXiv:2212.11387 [hep-lat]].

  
\bibitem{PoS:YHuo}  
``Fourier Acceleration of SU(3) Pure Gauge Theory at Weak Coupling''
  PoS(LATTICE2023)042 Y. Huo and N.H. Christ

\bibitem{Cossu:2017eys}
G.~Cossu, P.~Boyle, N.~Christ, C.~Jung, A.~J\"uttner and F.~Sanfilippo,
``Testing algorithms for critical slowing down,''
EPJ Web Conf. \textbf{175} (2018), 02008
doi:10.1051/epjconf/201817502008
[arXiv:1710.07036 [hep-lat]].

\bibitem{PoS:Jung}
``Riemannian manifold HMC with fermions''
  PoS(LATTICE2023)009 C. Jung

\bibitem{Boyle:2022pai}
P.~A.~Boyle, D.~Bollweg, C.~Kelly and A.~Yamaguchi,
``Algorithms for domain wall Fermions,''
PoS \textbf{LATTICE2021} (2022), 470
doi:10.22323/1.396.0470
[arXiv:2203.17119 [hep-lat]].


\bibitem{Duane:1988vr}
S.~Duane and B.~J.~Pendleton,
``GAUGE INVARIANT FOURIER ACCELERATION,''
Phys. Lett. B \textbf{206} (1988), 101-106
doi:10.1016/0370-2693(88)91270-1


\bibitem{Girolami}
  M. Girolami, B. Calderhead,
``Riemann Manifold Langevin and Hamiltonian Monte Carlo Methods ''
  Journal of the Royal Statistical Society: Series B (Statistical Methodology) 73, 123 (2011)



\bibitem{PoS:Tomiya}
  ``Equivariant transformer is all you need''
  PoS(LATTICE2023)001 A. Tomiya and Y. Nagai


\bibitem{PoS:DeSantis}
``Extraction of lattice QCD spectral densities from an ensemble of trained machines''
PoS(LATTICE2023)003 A. De Santis, M. Buzzicotti and N. Tantalo


\bibitem{PoS:Sheng}
  ``A Neural Network Approach to Lattice Field Theory''
  PoS(LATTICE2023)006 A. Sheng

\bibitem{PoS:Gantgen}
  ``Reducing the Sign Problem with simple Contour Deformation''
  PoS(LATTICE2023)007 C. G\"antgen

\bibitem{PoS:Orginos}
  ``Trivializing Flow in 2D O(3) sigma model''
  PoS(LATTICE2023)008 C. Chamness, K. Orginos and D. Kovner

  
\bibitem{PoS:Alvestad}
  ``Lattice real-time simulations with machine learned optimal kernels''
  PoS(LATTICE2023)010 D. Alvestad

\bibitem{PoS:Hackett}
  ``Practical applications of machine-learned flows on gauge fields''
  PoS(LATTICE2023)011 D. Hackett

\bibitem{PoS:Hoying}
  ``Unfreezing topology with nested sampling''
  PoS(LATTICE2023)012 D. Hoying

\bibitem{PoS:Rossney}  
``Learning Trivializing Flows in a $\phi^4$ theory from coarser lattices''
  PoS(LATTICE2023)013 attachments D. Albandea, L. Del Debbio, P. Hernandez, R. Kenway, J. Marsh Rossney and A. Ramos

\bibitem{PoS:Boyda}
    ``Enhancing Expressivity in Machine Learning: Application of Normalizing Flows in lattice QCD Simulations''
    PoS(LATTICE2023)014 D. Boyda

\bibitem{PoS:Caselle}
``Sampling Nambu-Goto theory using Normalizing Flows''
  PoS(LATTICE2023)015 M. Caselle, E. Cellini and A. Nada



\bibitem{PoS:Pederiva}  
``Scalar content of nucleon with the gradient flow using machine learning''
  PoS(LATTICE2023)019 G. Pederiva


  

\bibitem{PoS:Finkenrath}  
``Fine grinding localized updates via gauge equivariant flows in the 2D Schwinger model''
  PoS(LATTICE2023)022 J.F. Nieto Castellanos and J. Finkenrath

\bibitem{PoS:Osborn}  
``Tuning HMC parameters with gradients''
  PoS(LATTICE2023)023 J.C. Osborn



\bibitem{PoS:Urban}
``Constructing approximate semi-analytic and machine-learned trivializing maps for lattice gauge theory''
  PoS(LATTICE2023)026 J. Urban


  
  
  
\bibitem{PoS:Rodekamp}
``From Theory to Practice: Applying Neural Networks to Simulate Real Systems with Sign Problems''
  PoS(LATTICE2023)031 M. Rodekamp, E. Berkowitz, M. Dinca, C. Gantgen, S. Krieg and T. Luu




\bibitem{PoS:Abbott}
``Multiscale Normalizing Flows for Gauge Theories''
  PoS(LATTICE2023)035 R. Abbott

\bibitem{PoS:Foreman}  
``MLMC: Machine Learning Monte Carlo for Lattice Gauge Theory''
  PoS(LATTICE2023)036 S. Foreman, X.y. Jin and J.C. Osborn

  
\bibitem{PoS:Wettig}  
Gauge-equivariant multigrid neural networks
PoS(LATTICE2023)037 D. Kn\"uttel, C. Lehner and T. Wettig

\bibitem{PoS:Wenger}
Fixed point actions from convolutional neural networks
PoS(LATTICE2023)038 U. Wenger, K. Holland, A. Ipp and D.I. M\"uller



\bibitem{PoS:XYJin}
``Neural Network Gauge Field Transformation for 4D SU(3) gauge fields''
  PoS(LATTICE2023)040 X.Y. Jin

\bibitem{PoS:YLin}
``Signal-to-noise improvement through neural network contour deformations for 3D $SU(2)$ lattice gauge theory''
PoS(LATTICE2023)043 Y. Lin, W. Detmold, G. Kanwar, P. Shanahan and M. Wagman~




\bibitem{Boyle:2016lbp}
P.~A.~Boyle, G.~Cossu, A.~Yamaguchi and A.~Portelli,
``Grid: A next generation data parallel C++ QCD library,''
PoS \textbf{LATTICE2015} (2016), 023
doi:10.22323/1.251.0023

\bibitem{Yamaguchi:2022feu}
A.~Yamaguchi, P.~Boyle, G.~Cossu, G.~Filaci, C.~Lehner and A.~Portelli,
``Grid: OneCode and FourAPIs,''
PoS \textbf{LATTICE2021} (2022), 035
doi:10.22323/1.396.0035
[arXiv:2203.06777 [hep-lat]].


  
\bibitem{Swendsen:1986vqb}
R.~H.~Swendsen and J.~S.~Wang,
``Replica Monte Carlo Simulation of Spin-Glasses,''
Phys. Rev. Lett. \textbf{57} (1986) no.21, 2607
doi:10.1103/PhysRevLett.57.2607

\bibitem{Boyd:1997rb}
G.~Boyd,
``Tempered fermions in the hybrid Monte Carlo algorithm,''
Nucl. Phys. B Proc. Suppl. \textbf{60} (1998), 341-344
doi:10.1016/S0920-5632(97)00495-7
[arXiv:hep-lat/9712012 [hep-lat]].

\bibitem{Marinari:1992qd}
E.~Marinari and G.~Parisi,
``Simulated tempering: A New Monte Carlo scheme,''
EPL \textbf{19} (1992), 451-458
doi:10.1209/0295-5075/19/6/002
[arXiv:hep-lat/9205018 [hep-lat]].

\bibitem{Joo:1998ib}
B.~Joo \textit{et al.} [UKQCD],
``Parallel tempering in lattice QCD with O(a)-improved Wilson fermions,''
Phys. Rev. D \textbf{59} (1999), 114501
doi:10.1103/PhysRevD.59.114501
[arXiv:hep-lat/9810032 [hep-lat]].

\bibitem{Hasenbusch:2017unr}
M.~Hasenbusch,
``Fighting topological freezing in the two-dimensional CPN-1 model,''
Phys. Rev. D \textbf{96} (2017) no.5, 054504
doi:10.1103/PhysRevD.96.054504
[arXiv:1706.04443 [hep-lat]].

\bibitem{Bonanno:2022hmz}
C.~Bonanno,
``Lattice determination of the topological susceptibility slope \ensuremath{\chi}' of 2d CPN-1 models at large N,''
Phys. Rev. D \textbf{107} (2023) no.1, 014514
doi:10.1103/PhysRevD.107.014514
[arXiv:2212.02330 [hep-lat]].

\bibitem{Berni:2019bch}
M.~Berni, C.~Bonanno and M.~D'Elia,
``Large-$N$ expansion and $\theta$-dependence of $2d$ $CP^{N-1}$ models beyond the leading order,''
Phys. Rev. D \textbf{100} (2019) no.11, 114509
doi:10.1103/PhysRevD.100.114509
[arXiv:1911.03384 [hep-lat]].

\bibitem{Bonanno:2020hht}
C.~Bonanno, C.~Bonati and M.~D'Elia,
``Large-$N$ $SU(N)$ Yang-Mills theories with milder topological freezing,''
JHEP \textbf{03} (2021), 111
doi:10.1007/JHEP03(2021)111
[arXiv:2012.14000 [hep-lat]].

\bibitem{Bonanno:2022yjr}
C.~Bonanno, M.~D'Elia, B.~Lucini and D.~Vadacchino,
``Towards glueball masses of large-N SU(N) pure-gauge theories without topological freezing,''
Phys. Lett. B \textbf{833} (2022), 137281
doi:10.1016/j.physletb.2022.137281
[arXiv:2205.06190 [hep-lat]].

\bibitem{Bonanno:2022vot}
C.~Bonanno, M.~D'Elia, B.~Lucini and D.~Vadacchino,
``Towards glueball masses of large-$N$ $\mathrm{SU}(N)$ Yang-Mills theories without topological freezing via parallel tempering on boundary conditions,''
PoS \textbf{LATTICE2022} (2023), 392
doi:10.22323/1.430.0392
[arXiv:2210.07622 [hep-lat]].

\bibitem{PoS:DaSilva}
``The twisted gradient flow strong coupling with parallel tempering on boundary conditions''
PoS(LATTICE2023)354 J.L. Dasilva Golán, C. Bonanno, M. D’Elia, M. Garcia Perez and A. Giorgieri

\bibitem{Eichhorn:2023uge}
T.~Eichhorn, G.~Fuwa, C.~Hoelbling and L.~Varnhorst,
``Parallel Tempered Metadynamics: Overcoming potential barriers without surfing or tunneling,''
[arXiv:2307.04742 [hep-lat]].


\bibitem{PoS:Bonnano}
  ``Out-of-equilibrium simulations to fight topological freezing''
  PoS(LATTICE2023)005 C. Bonanno, A. Nada and D. Vadacchino

\bibitem{PoS:Matsumoto}
``Decimation map in 2D for accelerating HMC''
  PoS(LATTICE2023)033 N. Matsumoto, R.C. Brower and T. Izubuchi


%



\bibitem{Ce:2016idq}
M.~C\`e, L.~Giusti and S.~Schaefer,
Phys. Rev. D \textbf{93} (2016) no.9, 094507
doi:10.1103/PhysRevD.93.094507
[arXiv:1601.04587 [hep-lat]].

\bibitem{Ce:2016qto}
M.~C\`e, L.~Giusti and S.~Schaefer,
PoS \textbf{LATTICE2016} (2016), 263
doi:10.22323/1.256.0263
[arXiv:1612.06424 [hep-lat]].

%
\bibitem{Giusti:2018vxm}
L.~Giusti, T.~Harris, A.~Nada and S.~Schaefer,
PoS \textbf{LATTICE2018} (2018), 028
doi:10.22323/1.334.0028
[arXiv:1812.01875 [hep-lat]].

%
\bibitem{Giusti:2021qhk}
L.~Giusti, M.~D.~Brida, T.~Harris and M.~Pepe,
PoS \textbf{LATTICE2021} (2022), 356
doi:10.22323/1.396.0356
[arXiv:2112.02647 [hep-lat]].




\bibitem{Luscher:2017cjh}
M.~L\"uscher,
EPJ Web Conf. \textbf{175} (2018), 01002
doi:10.1051/epjconf/201817501002
[arXiv:1707.09758 [hep-lat]].

\bibitem{Giusti:2018cmp}
L.~Giusti and M.~L\"uscher,
Eur. Phys. J. C \textbf{79} (2019) no.3, 207
doi:10.1140/epjc/s10052-019-6706-7
[arXiv:1812.02062 [hep-lat]].

\bibitem{Fritzsch:2021klm}
P.~Fritzsch, J.~Bulava, M.~C\`e, A.~Francis, M.~L\"uscher and A.~Rago,
PoS \textbf{LATTICE2021} (2022), 465
doi:10.22323/1.396.0465
[arXiv:2111.11544 [hep-lat]].


\bibitem{Brezina}
``Adaptive Smoothed Aggregation ($\alpha$SA) Multigrid''
M. Brezina, R. Falgout, S. MacLachlan, T. Manteuffel, S. McCormick and J. Ruge
SIAM Review Vol. 47, No. 2 (Jun., 2005), pp. 317-346 (30 pages)


\bibitem{Trefethen}
N. M. Nachtigal, S. C. Reddy, and L. N. Trefethen, ”How Fast are Nonsymmetric Matrix Iterations?” SIAM. J. Matrix
Anal. Appl., 13(3), 778795.



\bibitem{Ishikawa:2021iqw}
K.~I.~Ishikawa, I.~Kanamori, H.~Matsufuru, I.~Miyoshi, Y.~Mukai, Y.~Nakamura, K.~Nitadori and M.~Tsuji,
``102 PFLOPS lattice QCD quark solver on Fugaku,''
Comput. Phys. Commun. \textbf{282} (2023), 108510
doi:10.1016/j.cpc.2022.108510
[arXiv:2109.10687 [hep-lat]].

\bibitem{Boku:2012zi}
T.~Boku, K.~I.~Ishikawa, Y.~Kuramashi, K.~Minami, Y.~Nakamura, F.~Shoji, D.~Takahashi, M.~Terai, A.~Ukawa and T.~Yoshie,
``Multi-block/multi-core SSOR preconditioner for the QCD quark solver for K computer,''
PoS \textbf{LATTICE2012} (2012), 188
doi:10.22323/1.164.0188
[arXiv:1210.7398 [hep-lat]].

\bibitem{PACS-CS:2009sof}
S.~Aoki \textit{et al.} [PACS-CS],
``Physical Point Simulation in 2+1 Flavor Lattice QCD,''
Phys. Rev. D \textbf{81} (2010), 074503
doi:10.1103/PhysRevD.81.074503
[arXiv:0911.2561 [hep-lat]].

\bibitem{PACS-CS:2011ngu}
Y.~Namekawa \textit{et al.} [PACS-CS],
``Charm quark system at the physical point of 2+1 flavor lattice QCD,''
Phys. Rev. D \textbf{84} (2011), 074505
doi:10.1103/PhysRevD.84.074505
[arXiv:1104.4600 [hep-lat]].

\bibitem{Ishikawa:2015rho}
K.~I.~Ishikawa \textit{et al.} [PACS],
``2+1 Flavor QCD Simulation on a $96^4$ Lattice,''
PoS \textbf{LATTICE2015} (2016), 075
doi:10.22323/1.251.0075
[arXiv:1511.09222 [hep-lat]].




\bibitem{PoS:Barca}
``Performance of two-level sampling for the glueball spectrum in pure gauge theory''
  PoS(LATTICE2023)030 L. Barca, F. Knechtli, M.J. Peardon, S. Schaefer and J.A. Urrea-Niño

\bibitem{PoS:XYu}  
``On the geometric convergence of HMC on Riemannian manifolds''
  PoS(LATTICE2023)041 X. Yu

\bibitem{PoS:Weinberg}
``HISQy Business''
  PoS(LATTICE2023)018 E. Weinberg

%
\bibitem{Lehner:2023bba}
C.~Lehner and T.~Wettig,
Phys. Rev. D \textbf{108} (2023) no.3, 034503
doi:10.1103/PhysRevD.108.034503
[arXiv:2302.05419 [hep-lat]].

%
\bibitem{Lehner:2023prf}
C.~Lehner and T.~Wettig,
[arXiv:2304.10438 [hep-lat]].
\end{thebibliography}
\end{document}